\documentclass[12pt]{article}
\usepackage{amssymb}
\usepackage{amsmath}
\usepackage{multirow}
\usepackage{latexsym}
\usepackage{epsfig}
\usepackage{graphicx}
\usepackage{epstopdf}
\usepackage[hang,small,tight]{subfigure} 
\usepackage[usenames,dvipsnames]{color}
\usepackage{here}

\def\boxit#1{\vbox{\hrule\hbox{\vrule\kern6pt
          \vbox{\kern6pt#1\kern6pt}\kern6pt\vrule}\hrule}}

\usepackage{natbib}

\newcommand{\bfh}{{\bf h}}

\newcommand{\bfs}{{\bf s}}

\newcommand{\bfy}{{\bf y}}

\newcommand{\bfY}{{\bf Y}}

\newtheorem{theorem}{Theorem}[section]
\newtheorem{proposition}{Proposition}[section]

\begin{document}
\thispagestyle{empty} \baselineskip=28pt

\begin{center}
{\LARGE{\bf Bayesian covariance modeling of multivariate spatial random fields}}
\end{center}

\baselineskip=12pt

\vskip 2mm
\begin{center}
Rafael S. Erbisti\footnote{\baselineskip=10pt Department of Statistics,
Federal University of Rio de Janeiro.} and
Thais C. O. Fonseca\footnote{\baselineskip=10pt  Department of Statistics,
Federal University of Rio de Janeiro. (To whom correspondence should be addressed; Department of Statistics, Federal University of Rio de Janeiro, Ilha do Fund\~ao, Av. Athos da Silveira Ramos, Centro de Tecnologia, Bloco C, CEP 21941-909, Rio de Janeiro, RJ, Brasil, thais@im.ufrj.br).} and
Mariane B. Alves\footnote{\baselineskip=10pt Department of Statistics,
Federal University of Rio de Janeiro.}
\end{center}
%
%
%
%

\begin{center}
{\large{\bf Abstract}}
\end{center}
\baselineskip=12pt

In this work we present full Bayesian inference for a new flexible nonseparable class of cross-covariance functions for multivariate spatial data. A Bayesian test is proposed for separability of covariance functions which is much more interpretable than parameters related to separability. Spatial models have been increasingly applied in several areas, such as environmental science, climate science and agriculture. These data are usually available in space, time and possibly for several processes. In this context the modeling of dependence is crucial for correct uncertainty quantification and reliable predictions. In particular, for multivariate spatial data we need to specify a valid cross-covariance function, which defines the dependence between the components of a response vector for all locations in the spatial domain. However, cross-covariance functions are not easily specified and the computational burden is a limitation for model complexity. In this work, we propose a nonseparable covariance function that is based on the convex combination of separable covariance functions and on latent dimensions representation of the vector components. The covariance structure proposed is valid and flexible. We simulate four different scenarios for different degrees of separability and compute the posterior probability of separability. It turns out that the posterior probability is much easier to interpret than actual model parameters. We illustrate our methodology with a weather dataset from Cear\'a, Brazil.

%
%
%

\baselineskip=12pt
\par\vfill\noindent

\par\medskip\noindent

\clearpage\pagebreak\newpage \pagenumbering{arabic}
\baselineskip=24pt
\section{Introduction}
Realistic modeling of multivariate data observed over space and time is of great interest in several application areas such as environmental science, climate science and agriculture. Often in geostatistical modeling, the data is considered a partial realization of a random function $Y(\bfs)$, $\bfs\in D\subseteq \Re^d$. Furthermore, in many applications, several quantities are measured for each location $\bfs$, resulting in a random vector $\mathbf{Y(s)}$, $\mathbf{Y(s)}\in \Re^{p}$. The main goal of this work is to contribute with realistic modeling of multivariate spatial data.

Although complexity in spatial models is a computational problem, some features have to be taken care in the realistic analysis of spatial data. Firstly, the spatial multivariate modeling of data is usually associated to the idea that data which are closer in space is more correlated than data further apart. Also, vector components are usually better predicted considering the component dependence of this vector. These general ideas are directly related to the cross-covariance function of spatial multivariate data, that is, $Cov(Y_j(\bfs),Y_{j'}(\bfs')$), $\bfs,\bfs'\in D\subseteq \Re^d$ which models the spatial dependence of $Y_j(.)$ and $Y_{j'}(.)$. The covariance functions considered need to be valid. Thus, construction of new realistic covariance functions usually rely on mathematical simplifications which are not necessarily followed by good fitting to data. 

An usual simplifying assumption in spatial data modeling is that the cross-covariance functions are separable. Separability implies that the covariance function for different processes and spatial locations can be computed as the product of a purely spatial covariance and a component covariance function. This might not be a realistic assumption for different processes across space, since it implies that, for two fixed locations $\bfs$ and $\bfs'$, the respective component covariance should be proportional. That is, when the spatial location varies, the covariance pattern for different components remains the same. \citet{CresHuang99} discusses some shortcomings of separable models in the context of spatiotemporal processes and point out that separable models are often chosen for convenience rather than for fitting the data well. \citet{Stein05} presents results about the limited kind of behaviours which these classes represent in practice. A consequence of the separability assumption is that the different $p$ processes will have the same spatial range, which is a very restrictive assumption.

Another restrictive assumption that is a consequence of separability is the symmetry of covariance functions. Separability implies full symmetry, thus a covariance function which is not symmetric is also nonseparable. In applied settings, symmetry is not realistic. For instance, processes which are influenced by air flows might have asymmetric covariance functions.

Several authors have proposed models to relax the separability assumption of cross-covariance functions. The linear model of coregionalization  defines that the spatial process $\{Y_j(\bfs); j=1, \ldots, p\}$ can be decomposed into sets $\{W_u^{(j)}(\bfs); u=0,\ldots, K\}$ of spatially uncorrelated components, i.e, $Y_j(\bfs)=\sum_{u=0}^{K}W_u^{(j)}(\bfs)$, $u=0,\ldots, K$. In this approach, the cross-covariance functions $C_{ij}^{u}(\bfh)$ associated with the spatial components are composed of real coefficients $b_{ij}^{u}$ and are proportional to real correlation functions $\rho^{u}(\bfh)$, that is, $C_{ij}(\bfh)= \sum_{u=0}^{K} C_{ij}^{u}(\bfh) = \sum_{u=0}^{K} b_{ij}^{u}\rho^{u}(\bfh)$, with $\bfh$ the spatial separation vector \citep{Wack98}.

In a recent paper, \citet{Cres15} proposed the conditional approach to derive multivariate models. The construction is based on partitioning the vector of spatial processes so that the joint distribution is specified through univariate spatial conditional distributions. This is convenient as the modeler just needs to specify univariate covariance functions and an integrable function of $p$ arguments. Obviously, the results will depend on the chosen conditioning and this is not always an easy modeling decision.

A different proposal considers multidimensional scaling ideas \citep{Cox00}. Following this idea, \cite{Gent10} proposed a multivariate spatiotemporal model based on latent dimensions and distances between components. The authors represent the vector of components as coordinates in a $k-$dimensional space. Any valid covariance function can be used considering the latent component distances and spatial distances to define cross-covariances. Moreover, the authors present results of a simulated study where the model compares favorably to the coregionalization set-up which seems to lack flexibility for some scenarios. The approach of \cite{Gent10} depends on the specification of nonseparable covariance functions. In the paper they considered the function proposed in \cite{Gneit02}. The functions presented in \cite{Gneit02} are not interpretable or intuitive and the range of nonseparability achieved is limited \citep{FonsSteel17}. In this paper, we follow the multidimensional scaling approach and consider an interpretable class of nonseparable covariance functions.

In that context, this work extends the class of nonseparable covariance functions proposed in \cite{Fons10} to the modeling of component and spatial dependence and considers the multidimensional scaling ideas to define latent distances between components as in \citet{Gent10}. The general proposed class is able to model different ranges in space and asymmetric covariance structures. Furthermore, the proposed class allows for different degrees of smoothness across space for different components of the multivariate random vector. Also, the proposed class has subclasses which can possess a covariance function with the same differentiability properties as the Mat\'ern class. Similarly to the conditional approach of \citet{Cres15}, the proposed covariance depends on the definition of univariate covariances and a bivariate joint density function. It is advantageous compared to the conditional approach as it depends on a bivariate density function even if $p$ is large. In particular, the bivariate functions used in our proposal are trivially defined in terms of moment generating functions of univariate random variables, while the $b$-functions of Cressie and Zammit-Mangion are not easily interpretable. 

The remainder of the paper is organized as follows. Section \ref{section2} presents definitions and characteristics about multivariate process modeling. A new class of multivariate spatial covariances is presented in Section \ref{section3}. Inference on these models will be conducted from a Bayesian perspective and will be described in Section \ref{sec_inference}. Section \ref{section7} develops a Bayesian test of separability to measure the level of separability between space and components.  Simulated examples are presented in Section \ref{simulation}. Section \ref{application} presents an illustration of the proposed approach with a weather dataset. Finally, Section \ref{conclusions} presents conclusions and future developments.

\section{Multivariate process modeling} \label{section2}

In the context of multivariate spatial processes, the main goal is usually to model the dependence among several variables measured across a spatial domain of interest, in order to obtain realistic predictions. Denote by $\bf{Y}(\bf{s})$ the $p-$dimensional vector of variables at location $\textbf{s}\in D$. Thus, the direct covariance function measures the spatial dependence for each component individually, while the cross-covariance function between two random functions measures the component dependence at the same location and the component dependence within two different locations.

Assuming that $\textbf{Y}(\textbf{s})$ is a spatially stationary process, that is
\begin{equation}
E[Y_{i}(\textbf{s})]=m_i, \hspace{0.5cm}
Cov[Y_i(\textbf{s}),Y_j(\textbf{s}+\textbf{h})]=C_{ij}(\textbf{h}), \hspace{0.5cm}  \forall \textbf{s}, \textbf{s}+\textbf{h}\in D; i,j=1, 2,\ldots, p, \nonumber
\end{equation}
the cross-covariance function of $\textbf{Y}(\textbf{s})$ is defined as
\begin{equation} \label{crossC}
E[(Y_{i}(\textbf{s})-m_i)(Y_{j}(\textbf{s}+\textbf{h})-m_j)]=C_{ij}(\textbf{h}), \hspace{0.5cm} \textbf{s},\textbf{s}+\textbf{h} \in D; i,j=1, 2,\ldots, p.
\end{equation}

The requirement of positive definiteness of $C_{ij}(\cdot)$ is a limitation in the definition of realistic covariance functions for multivariate spatial processes. As a result, several simplifications are called for in practice such as stationarity and separability. Separability states that
\begin{equation}
C_{ij}(\bfs,\bfs')=a_{ij}\rho(\bfs,\bfs'),
\label{eq:separab}
\end{equation}
with $\boldsymbol{A}=\{a_{ij}\}$ a positive definite $p\times p$ matrix and $\rho(\cdot,\cdot)$ a valid correlation function. Let $\bfY$ be a vectorized version of $Y_{ik}=Y_i(\bfs_k)$, $k =1,\cdots,n; i=1,\cdots,p$. Then the covariance matrix is $\boldsymbol{\Sigma}=\bf{R}\otimes \bf{A}$, with $R_{kl}=\rho (\bfs_k,\bfs_l),$ $k,l =1,\cdots,n$. The condition of positive definiteness is respected if $\bf{R}$ and $\bf{A}$ are positive definite. This specification is computationally advantageous as inverses and determinants are obtained from smaller matrices, that is, $\boldsymbol{\Sigma}^{-1} = \textbf{R}^{-1} \otimes \textbf{A}^{-1}$ and $|\boldsymbol{\Sigma}| = |\textbf{R}|^{p} |\textbf{A}|^{n}$. However, this model has theoretical limitations \citep{GelfB03}. Firstly, it is an intrinsic model implying that the correlation between two components $Y_i(\bfs_k)$ and $Y_j(\bfs_l)$ is $a_{ij}$, that is, it does not depend on the locations $\bfs_k$ and $\bfs_l$. Secondly, note that as the covariance is defined by one spatial correlation function $\rho(\cdot,\cdot)$, the spatial range will be the same for all components. This last feature can be perceived through the following argument: consider the univariate spatial processes $\{Y(\textbf{s}): \textbf{s} \in D\}$ and $\{X(\textbf{s}): \textbf{s} \in D\}$,  $D \subset \Re^{2}$, therefore $\textbf{Y} = [Y(\textbf{s}_1), Y(\textbf{s}_2),\ldots, Y(\textbf{s}_n)]^{T}$ and  $\textbf{X} = [X(\textbf{s}_1), X(\textbf{s}_2),\ldots, X(\textbf{s}_n)]^{T}$.

It is possible to express the following linear relationship for any point in $D$: 
\begin{equation} \label{eq:relacaolinear}
E[Y| X]=\beta_{0}+\beta_{1}X.
\end{equation}

Consider the stacked $2n \times 1$ vector $(\textbf{X},\textbf{Y})^{T}$, following a multivariate Normal distribution and a separable covariance structure as in (\ref{eq:separab}), that is,
\begin{displaymath}
\left(\begin{array}{c}
\textbf{X}\\
\textbf{Y}\\
\end{array}\right)
\sim N_{2n}(\boldsymbol{\mu},\boldsymbol{\Sigma}), \hspace{0.5cm}\boldsymbol{\Sigma} = \textbf{A} \otimes \textbf{R},
\end{displaymath}
implying that $\textbf{X} \sim N_{n}(\boldsymbol{\mu}_{x}, a_{11}\textbf{R})$ and $\textbf{Y} \sim N_{n}(\boldsymbol{\mu}_{y}, a_{22}\textbf{R})$. It follows directly that $\textbf{Y}|\textbf{X} \sim N_{n}(\boldsymbol{\mu}^{*}, \boldsymbol{\Sigma}^{*})$, with
$
\boldsymbol{\mu}^{*} = \boldsymbol{\mu}_{y} - \frac{a_{12}}{a_{11}} \boldsymbol{\mu}_{x} + \frac{a_{12}}{a_{11}} \textbf{X}   $
 and
$\boldsymbol{\Sigma}^{*} =\left(a_{22} - \frac{a_{12}^{2}}{a_{11}}\right) \textbf{R}, $
which is equivalent to  $\textbf{Y}|\textbf{X} \sim N_{n}(\beta_{0} + \beta_{1}\textbf{X},  \sigma^{2}\textbf{R})$, with
$\beta_{0} = \boldsymbol{\mu}_{y} - \frac{a_{12}}{a_{11}} \boldsymbol{\mu}_{x}$, 
$\beta_{1} = \frac{a_{12}}{a_{11}}$ and
$\sigma^{2} = a_{22}-\frac{a_{12}^{2}}{a_{11}}.$

If we assume, reversely,  $\textbf{X} \sim N_{n}(\boldsymbol{\mu}_{x},a_{11}\textbf{R})$ and $\textbf{Y}|\textbf{X} \sim N_{n}(\beta_{0}+\beta_{1}\textbf{X}, \sigma^{2}\textbf{S})$, with \textbf{S} any spatial correlation matrix, the covariance structure for \textbf{Y} is 
\begin{align}\label{eq:cov_y_sep}
Cov[Y_{i}, Y_{j}] & = \sigma^{2} \textbf{S}_{ij}+\beta_{1}^{2} a_{11} \textbf{R}_{ij} \nonumber \\
& = a_{22} \textbf{S}_{ij} - \frac{a_{12}^{2}}{a_{11}} \textbf{S}_{ij} + \frac{a_{12}^{2}}{a_{11}} \textbf{R}_{ij}.
\end{align}
Then  (\ref{eq:cov_y_sep}) equals $a_{22}\textbf{R}$, reducing to the separable specification if and only if $\textbf{S}=\textbf{R}$, that is, if $\textbf{Y}|\textbf{X}$ has the same spatial correlation structure as \textbf{X}.

More flexible structures are obtained via the coregionalization approach, which in its simplest form is
 $\textbf{Y}(\textbf{s})=\textbf{Aw}(\textbf{s})$, with $\textbf{A}$ a \textit{p}$\times$\textit{p} matrix and the components of $\textbf{w}(\textbf{s})$, $w_{j}(\textbf{s})$, $j=1, 2,\ldots, p$, independent and identically distributed spatial processes. If the processes $w_{j}(\textbf{s})$ are stationary with zero mean and unit variances and $Cov(w_{j}(\textbf{s}),w_{j}(\textbf{s}'))=\rho(\textbf{s}-\textbf{s}')$, then $E(\textbf{Y}(\textbf{s}))=0$ and the cross-covariance function of $\textbf{Y}(\textbf{s})$ is
$
\boldsymbol{\Sigma}_{\textbf{Y}(\textbf{s}),\textbf{Y}(\textbf{s}')} \equiv C(\textbf{s}-\textbf{s}') = \rho(\textbf{s}-\textbf{s}') \textbf{AA}^{T}
$
which is separable. A more general form for the coregionalization model considers independent processes $w_{j}(\textbf{s})$ however they are not identically distributed. The covariance matrix is given by
\begin{displaymath}
\boldsymbol{\Sigma}_{\textbf{Y}(\textbf{s}),\textbf{Y}(\textbf{s}')} \equiv C(\textbf{s}-\textbf{s}') = \sum_{j=1}^{p} \rho_{j}(\textbf{s}-\textbf{s}') \textbf{T}_{j}
\end{displaymath}
with $\textbf{T}_{j}=\textbf{a}_{j}\textbf{a}_{j}^{T}$, $\textbf{a}_{j}$ the $j-th$ column of $\textbf{A}$. The resulting covariance is nonseparable but is stationary.

In this work we follow the multidimensional scaling framework and the latent dimensions proposed in \cite{Gent10}. The vector of components are represented as coordinates in a $k-$dimensional space, for an integer $1 \leq k \leq p$, that is, the $i-th$ component is represented as $\xi_{i}=(\xi_{i1},\ldots, \xi_{ik})^{T}$.

This approach can be used for any valid covariance function $\Sigma_{ij}=C((\textbf{s},\xi_{i}),(\textbf{s}',\xi_{j}))$. For any $\textbf{s}$, $\textbf{s}'$ there is $C_{\textbf{s},\textbf{s}'}(.)$ such that $C_{ij} (\textbf{s},\textbf{s}')=C_{\textbf{s},\textbf{s}'} (\xi_{i},\xi_{j})$ for some $\xi_{i}$, $\xi_{j} \in \Re^{k}$. A review of the main approaches to building a valid multivariate cross-covariance function is presented in \citet{Gent15}.

The latent coordinates may be treated as parameters and estimated from data. Moreover, it is possible to consider the reparametrisation $\delta_{ij}=\|\xi_{i}-\xi_{j}\|$. This approach is similar to the multidimensional scaling \citep{Cox00} with latent distances $\delta_{ij}$'s, where for fixed locations $\textbf{s}$ and $\textbf{s}'$, small $\delta_{ij}$'s are converted into strong cross-correlation. Notice that large values of $\delta_{ij}$'s mean small correlation. A further discussion about this issue is presented in the conclusions. 

As follows we consider an intuitive proposal for the construction of nonseparable covariance structures, which is based on mixing separable functions as in \cite{Fons10}.

\section{Multivariate spatial modeling based on mixtures} \label{section3}

In this section we present a new class of multivariate spatial covariances which are flexible and intuitive depending only on the specification of univariate functions on space. We consider the latent dimension approach of \cite{Gent10} to model cross-dependencies between components of a spatial vector. Furthermore we define the nonseparable function based on the spatiotemporal mixture approach of \cite{Fons10}. Thus, only univariate valid spatial functions need to be specified.

\cite{Fons10} consider $(\bfs,t) \in D \times T$, $D \subseteq \Re^{d}$, $T \subseteq \Re$, as space-time coordinates varying continuously on $D \times T$ and $Z_{1}(\bfs)$, $Z_{2}(t)$ uncorrelated processes, $\{Z_{1}(\bfs): \bfs \in D\}$ denoting a purely spatial process with covariance $C(\bfs)$ and $\{Z_{2}(t): t \in T\}$ a purely temporal process with covariance $C(t)$. The mixture representation of the covariance structure of $Z(\bfs,t)$ is defined as follows: assume that $(U,V)$ is a nonnegative bivariate random vector following a joint distribution $G(u,v)$, independent of $\{Z_{1}(\bfs): \bfs \in D\}$ and $\{Z_{2}(t): t \in T\}$. Define the process $Z(\bfs,t)= Z_{1}(\bfs,U)Z_{2}(t,V)$, where $\{Z_{1}(\bfs,u)\}$ remains a purely spatial process for every $u \in \Re_+$ with a stationary covariance function $C(\bfs,u)$ for $\bfs \in D$ and every $u \in \Re_+$, which is a  measurable function of $u \in \Re_+$ for every $\bfs \in D$. Analogously, let $\{Z_{2}(t,v)\}$ be a purely temporal process with covariance $C(t,v)$, which is a stationary covariance function for $t \in T$ and every $v \in \Re_+$ and a measurable function of $v \in \Re_+$ for every $t \in T$. Thus the corresponding covariance of $Z(\bfs,t)$ is a convex combination of separable covariance functions. This is a valid and generally nonseparable function
\begin{equation}\label{eq:mistura_fonseca}
C(\textbf{s},t)=\int \int C(\textbf{s};u) C(t;v) g(u,v)du dv.
\end{equation}

The proposed idea in the present work is to modify (\ref{eq:mistura_fonseca}) to deal with the multivariate spatial specification. Thus, consider $(U,V)$ independent of the process $\textbf{Y}(\textbf{s})$. Similar to \cite{Fons10}, the covariance of $\textbf{Y}(\textbf{s})$ is a convex combination of separable covariance functions, given by
\begin{equation}\label{eq:mistura}
C_{ij}(\textbf{s},\xi)=\int \int C(\textbf{s};u) C_{ij}(\xi;v) g(u,v)du dv
\end{equation}
with $\xi$ representing a latent dimension as in \cite{Gent10} and \textbf{s} an arbitrary spatial location. 

According to \cite{Fons10}, the fundamental step in the definition of this class of functions lies on the representation of the dependence between $U$ and $V$. Define variograms $\gamma_{1}(\textbf{s})$ and $\gamma_{2}(\xi)$ as continuous functions on $\textbf{s} \in \Re^{d}$ and $\xi \in \Re^{k}$, respectively. Then, it is possible to analytically solve (\ref{eq:mistura}), still assuring that  the generated covariance is positive definite, defining $C(\textbf{s};u)=exp\{-\gamma_{1}(\textbf{s})u\}$ and $C(\xi;v)=exp\{-\gamma_{2}(\xi)v\}$. 

\begin{proposition}\label{eq:prop1}
Consider a bivariate nonnegative vector $(U,V)$ with joint moment generator function $M(.,.)$. If the variograms $\gamma_{1}(\textbf{s})$ and $\gamma_{2}(\xi)$ are continuous functions of $\textbf{s} \in \Re^{d}$ and $\xi \in \Re^{k}$, respectively, and $C(\textbf{s};u)=exp\{-\gamma_{1}(\textbf{s})u\}$, $C(\xi;v)=exp\{-\gamma_{2}(\xi)v\}$, then (\ref{eq:mistura}) implies that
\begin{equation}
\rho_{ij}(\textbf{s},\xi)=M(-\gamma_{1}(\textbf{s}),-\gamma_{2}(\xi)),
\end{equation}
which is a valid correlation function.
\end{proposition}

\cite{Ge07m} use Monte Carlo integration to solve an integral similar to (\ref{eq:mistura}). \cite{Apana12} consider a multivariate version of  Mat\'ern, presenting a flexible model, allowing for different behaviour for each component. The proposed approach (\ref{eq:mistura}) also presents that flexibility.

Following Proposition \ref{eq:prop1}, it is possible to build nonseparable structures, based only on the joint distribution of $(U,V)$. Thus, consider the following proposition.

\begin{proposition}\label{eq:prop2}
Consider the independent nonnegative random variables $X_{0}$, $X_{1}$ and $X_{2}$, with moment generator functions $M_{0}$, $M_{1}$ and $M_{2}$. Define  $U$ and $V$ as: $U=X_{0}+X_{1}$ and $V=X_{0}+X_{2}$. If $C(\textbf{s};u)=exp\{-\gamma_{1}(\textbf{s})u\}$ and $C(\xi;v)=exp\{-\gamma_{2}(\xi)v\}$, as in proposition \ref{eq:prop1}, then the correlation function resulting from (\ref{eq:mistura}) is
\begin{equation}\label{eq:correlacao}
\rho_{ij}(\textbf{s},\xi)=M_{0}(-\gamma_{1}(\textbf{s})-\gamma_{2}(\xi))M_{1}(-\gamma_{1}(\textbf{s}))M_{2}(-\gamma_{2}(\xi)).
\end{equation}

\end{proposition}

Observe that if $U$ and $V$ are uncorrelated, that is, $U=X_{1}$ and $V=X_{2}$, the separable specification is obtained, since $\rho_{ij}(\textbf{s},\xi)=M_{1}(-\gamma_{1}(\textbf{s}))M_{2}(-\gamma_{2}(\xi))$. 

The class generated by proposition \ref{eq:prop2} allows for different parametric representations, as we vary the specifications for $X_0$, $X_1$ and $X_2$. By construction, any non-null correlation between $U$ and $V$ will be positive.

Proposition \ref{eq:prop2} generates a valid correlation structure, thus, following \cite{Ge07m}, a valid covariance structure is given by
\begin{equation}\label{eq:validando}
\rho_{ij}(\textbf{s},\xi)=\frac{C_{ij}(\textbf{s},\xi)}{[C_{ii}(\textbf{0})C_{jj}(\textbf{0})]^{1/2}}.
\end{equation}

Note that $\rho_{ii}(\textbf{0})=1$. Consider a diagonal matrix $D_{cov}$ with elements $[D_{cov}]_{ii}=C_{ii}(\textbf{0})$. If $R(\textbf{s},\xi)=D_{cov}^{-1/2} C(\textbf{s},\xi) D_{cov}^{-1/2}$, then $R(\textbf{s},\xi)$ is a valid cross-correlation matrix. If we define $D_{\sigma}^{1/2}=diag(\sigma_{1},\ldots,\sigma_{p})$, $\sigma_{i} \in \Re$, $i = 1,\ldots,p$, a valid cross-covariance structure is obtained, given by the matrix $C(\textbf{s},\xi)=D_{\sigma}^{1/2}R(\textbf{s},\xi)D_{\sigma}^{1/2}$.

\begin{proposition}\label{eq:prop3}
Consider the nonnegative independent variables $X_{0}$, $X_{1}$ and $X_{2}$, with moment generator functions $M_{0}$, $M_{1}$ and $M_{2}$. Define $U$ and $V$ as: $U=X_{0}+X_{1}$, $V=X_{0}+X_{2}$. If $C(\textbf{s};u)=exp\{-\gamma_{1}(\textbf{s})u\}$ and $C(\xi;v)=exp\{-\gamma_{2}(\xi)v\}$, then (\ref{eq:mistura}) and (\ref{eq:validando}) imply that
\begin{equation}\label{eq:class_correlacao}
C_{ij}(\textbf{s},\xi)=\sigma_{i} \sigma_{j} M_{0}(-\gamma_{1}(\textbf{s})-\gamma_{2}(\xi))M_{1}(-\gamma_{1}(\textbf{s}))M_{2}(-\gamma_{2}(\xi)),
\end{equation}
which is a valid covariance function.
\end{proposition}

\subsection{Flexible classes} \label{sec_flexible}

In this section we present a new covariance function following proposition \ref{eq:prop3}. Consider $X_{0}$, $X_{1}$ and $X_{2}$ following gamma distributions as \cite{Fons10}. 	
\begin{theorem}\label{eq:teo1}
Consider $X_{l} \sim Gamma(\alpha_{l},\lambda_{l})$, $l=0, 1$ and $2$, and the variograms $\gamma_{1}(\textbf{s})$ and $\gamma_{2}(\xi)$ as continuous functions of $\textbf{s} \in \Re^{d}$ and $\xi \in \Re^{k}$, respectively, then from proposition \ref{eq:prop3}, the cross-covariance function is
\begin{equation}\label{eq:func_prop}
C_{ij}(\textbf{s},\xi)=\sigma_{i} \sigma_{j} \left(1+\frac{\gamma_{1}(\textbf{s})+\gamma_{2}(\xi)}{\lambda_{0}}\right)^{-\alpha_{0}} \left(1+\frac{\gamma_{1}(\textbf{s})}{\lambda_{1}}\right)^{-\alpha_{1}} \left(1+\frac{\gamma_{2}(\xi)}{\lambda_{2}}\right)^{-\alpha_{2}} 
\end{equation}
with $\sigma_{k}\in \Re$, $k=1,\ldots, p$, $\alpha_{l}>0$ and $\lambda_{l}>0$, $l=0, 1, 2$.
\end{theorem}

It is difficult to interpret some parameters in the proposed function (\ref{eq:func_prop}). 
We expect to work a function that allows different spatial ranges for each component. The dependence between $U$ and $V$ is governed by the variable $X_{0}$, it is important to define a parameter responsible for the behaviour of the correlation between these variables. Remember that if $U$ and $V$ are uncorrelated then the separable case is obtained. 

In order to achieve those goals and to avoid redundancy, like \cite{CresHuang99} and \cite{Fons10}, we fix $\lambda_{i} = 1$, for $i=0, 1$ e $2$, and work with a component variogram $\gamma_{2}(\xi)=\|\xi_{i}-\xi_{j}\|=\delta_{ij}$ . Furthermore, we introduce an extra parameter in the spatial variogram allowing for different spatial ranges. This parameter varies with the components $i$ and $j$, i.e, $\gamma_{1}(\textbf{s})=\frac{\| \textbf{s}-\textbf{s}' \|}{b_{ij}}=\frac{h}{b_{ij}}$. Therefore, the general model is given by
\begin{equation}\label{eq:func_prop_repa}
C_{ij}(\textbf{s},\xi)=\sigma_{i} \sigma_{j} \left(1+\delta_{ij}+\frac{h}{b_{ij}}\right)^{-\alpha_{0}} \left(1+\frac{h}{b_{ij}}\right)^{-\alpha_{1}} \left(1+\delta_{ij}\right)^{-\alpha_{2}} 
\end{equation}
where $\delta_{ij}$ is the latent distance between the components $i$ and $j$, $\sigma_{i}$ is the covariance of component $i$, $b_{ij}$'s are spatial range parameters, $\alpha_{l}$ are smoothness parameters, for $l=1$ and $2$, and $\alpha_{0}$ is a separability parameter.

Notice that if we work with the same spatial range parameters for all components, that it, $b_{ij} = \phi$, $\forall i,j= 1, 2,\ldots, p$, we provide a particular case of the general function. Furthermore, if $\alpha_{0}=0$, the separable model is obtained and the resulting covariance function is in the Cauchy class. However, the general class is flexible enough to generate a nonseparable covariance structure and allows for different spatial ranges associated to each component.

\section{Bayesian inference} \label{sec_inference}

Let $(\textbf{y}_{t}(\textbf{s}_1),\ldots,\textbf{y}_{t}(\textbf{s}_n))$ be a matrix of multivariate data observed at spatial locations $\bfs_1,\ldots,\bfs_n \in D$ and at time $t$, where $\textbf{y}_{t}(\textbf{s}_i)=(y_{1t}(\textbf{s}_i),\ldots, y_{pt}(\textbf{s}_i))'$, $t=1,\ldots,T$, is a $p-$dimensional vector. If the Gaussian assumption is made, the likelihood function with $T$ independent replicates for the unknown parameters based on $n$ spatial locations is given by
\begin{align}\label{eq:verossimilhanca}
l(\textbf{y};\boldsymbol{\theta}) & = \prod_{t=1}^{T} (2\pi)^{\frac{-np}{2}}|\boldsymbol{\Sigma}|^{-1/2}exp\left\{\frac{-1}{2}(\textbf{y}_t-\boldsymbol{\mu})'\boldsymbol{\Sigma}^{-1}(\textbf{y}_t-\boldsymbol{\mu})\right\} \nonumber \\
& = (2\pi)^{\frac{-npT}{2}}|\boldsymbol{\Sigma}|^{-T/2} exp\left\{\frac{-1}{2} \sum_{t=1}^{T} (\textbf{y}_t-\boldsymbol{\mu})'\boldsymbol{\Sigma}^{-1}(\textbf{y}_t-\boldsymbol{\mu})\right\}
\end{align}
with $\textbf{y}_{t}$ the vectorized version of $(\textbf{y}_{t}(\textbf{s}_1),\ldots,\textbf{y}_{t}(\textbf{s}_n))$ with $np$ observations, $\boldsymbol{\mu}=\textbf{X}\boldsymbol{\beta}$ the mean vector, $\boldsymbol{\Sigma}$ the covariance matrix with dimension $np \times np$, and $\boldsymbol{\theta}$ the parameter vector. The covariance matrix has components defined by equation (\ref{eq:func_prop_repa}). In particular for our model specification $\boldsymbol{\theta} = (\boldsymbol{\sigma}, \boldsymbol{\delta}, \boldsymbol{\alpha}, \textbf{b},\boldsymbol{\beta})$, with $\boldsymbol{\sigma}=(\sigma_{1},\ldots, \sigma_{p})$, $\boldsymbol{\delta}$ the vector of latent variables $\delta_{ij}$, $i \neq j$, $i,j = 1,\ldots, p$, $\boldsymbol{\alpha}=(\alpha_{0}, \alpha_{1}, \alpha_{2})$, \textbf{b} the range parameter vector $b_{ij}$, $i,j = 1,\ldots, p$, and $\boldsymbol{\beta}=(\beta_{10},\ldots, \beta_{p0},\beta_{11},\ldots,\beta_{p1},\ldots,$ $\beta_{1q},\ldots, \beta_{pq})$, with $q$ the number of covariates.

To complete the Bayesian model specification, the prior distributions must be defined for all parameters in the proposed model (\ref{eq:func_prop_repa}).
Prior independence is assumed for the parameters in the model such that $\sigma_{i} \sim N(c_{i},d_{i})$, $i = 1,\ldots, p$, $\delta_{ij} \sim Ga(f_{ij},g_{ij})$,  $i \neq j$, $i,j = 1,\dots, p$, $\alpha_{k} \sim Ga(r_{k},s_{k})$, $k= 0, 1, 2$, $b_{ij} \sim Ga(u_{ij}\times med(d_{s}),u_{ij})$, $i,j = 1,\ldots, p$, $med(d_{s})$ denoting the median of the spatial distances, $\boldsymbol{\beta} \sim N_{pq}(\boldsymbol{\lambda},\boldsymbol{\Lambda})$.

Inference is based on simulations from the complete conditional distributions for sets of parameters. The complete conditional distribution for $\boldsymbol{\beta}$ is Gaussian. For the other parameters in the covariance function the distributions have no closed form and Metropolis-Hastings steps are considered in the Gibbs sampler algorithm. Details on such algorithms are presented in \citet{danilopes}. 

\subsection{Prediction}

One of the main goals in spatial data analysis is to obtain prediction in new locations or for missing data within the observed data. Let $\textbf{y}_{u}$ be the observation vector at unmeasured locations $\textbf{s}_{u} \in D$. The prediction of $\textbf{y}_{u}$ is based on the predictive distribution $p(\textbf{y}_{u}|\textbf{y}_{o})$, with $\textbf{y}_{o}$ denoting the vector of observed data. Thus,
\begin{align}\label{eq:predict_dist}
p(\textbf{y}_{u}|\textbf{y}_{o}) & =  \int p(\textbf{y}_{u}|\textbf{y}_{o},\boldsymbol{\theta}) p(\boldsymbol{\theta}|\textbf{y}_{o}) d\boldsymbol{\theta}.
\end{align}

From the Gaussian assumption, the distribution $p(\textbf{y}_{u}|\textbf{y}_{o},\boldsymbol{\theta})$ is also Gaussian with parameters
$
\boldsymbol{\mu}^{*} = \boldsymbol{\mu}_{u}+\boldsymbol{\Sigma}_{uo}\boldsymbol{\Sigma}^{-1}_{oo}(\textbf{y}_{o}-\boldsymbol{\mu}_{o})
$
and
$
\boldsymbol{\Sigma}^{*} = \boldsymbol{\Sigma}_{uu} - \boldsymbol{\Sigma}_{uo}\boldsymbol{\Sigma}^{-1}_{oo}\boldsymbol{\Sigma}_{ou}.
$
Assume that $\boldsymbol{\theta}^{(1)},\ldots, \boldsymbol{\theta}^{(M)}$ are a sample from the posterior distribution $(\boldsymbol{\theta}|\textbf{y}_{o})$ obtained by MCMC sampling. Thus, the predictive distribution in (\ref{eq:predict_dist}) may be obtained by the approximation:
\begin{equation}
\widehat{p}(\textbf{y}_{u}|\textbf{y}_{o}) = \frac{1}{M} \sum_{i=1}^{M} p(\textbf{y}_{u}|\textbf{y}_{o},\boldsymbol{\theta}^{(i)}).
\end{equation}

\section{Bayesian hypotheses testing for separability} \label{section7}

Following \cite{Fons10}, we choose the correlation between $U$ and $V$ as a measure of separability. Indeed, if $U$ and $V$ are uncorrelated, the resulting model is separable, so
\begin{align}
\tilde\rho = Corr(U,V) & = \frac{Cov(U,V)}{\sqrt[]{Var(U) Var(V)}} \nonumber \\
& = \frac{\alpha_{0}}{\sqrt[]{(\alpha_{0}+ \alpha_{1}) (\alpha_{0}+\alpha_{2})}}.  \nonumber
\end{align}

It is easy to see that $\alpha_{0} = 0$ implies $\tilde \rho = 0$. Note that $0 \leq \tilde \rho \leq 1$, where 0 indicates separability and 1 indicates strong nonseparability. From a frequentist point of view, many authors present a formal method to test separability in the spatiotemporal models \citep{Mit05, Mit06, Fuentes06a}. The test proposed in this work aims to measure the degree of separability between space and components and we follow the Bayesian paradigm for hypothesis testing.

\subsection{Bayesian model choice}

The usual continuous prior for positive parameters, as the one considered for $\alpha_0$ in Section \ref{sec_inference}, assigns zero probability for the null hypothesis $\alpha_0=0$. As an alternative consider the following mixture representation
\begin{equation} \label{prior_mixture}
\pi(\alpha_0)=p_0\mathcal{D}_{0}+(1-p_0)g(\alpha_0),
\end{equation}
with $\mathcal{D}_0$ the dirac function at $\alpha_0=0$ and $g(\alpha_0)$ a continuous distribution for $\alpha_0>0$. Thus, $p_0$ is the prior probability of a separable covariance function. The resulting posterior distribution in this specification is also a mixture
$$\pi(\alpha_0\mid \bfy)=\tilde p_0\mathcal{D}_{0}+(1-\tilde p_0)g(\alpha_0\mid \bfy),$$
with $\tilde p_0$ being the posterior probability of separable covariance functions given the data.

The posterior probabilities $\tilde p_0$ might be used to select a model (Bayesian model choice) or to predict new observations based on model averaging across both models \citep{hoeting99}.

Therefore, consider a general situation in which it is desired to test the null hypothesis $H_0$: $\alpha_0 \in \Omega_0$ and the alternative hypothesis $H_1$: $\alpha_0 \in \Omega_1$, where $\Omega_0 \cup \Omega_1$ is the entire parameter space.
Let $d_0$ be the decision of not rejecting the null hypothesis $H_0$ and let $d_1$ be the decision of rejecting $H_0$. We can assume a loss $w_0$ by taking decision $d_1$ when $H_0$ is true, and a loss $w_1$ if we take decision $d_0$ when $H_1$ is true. The general idea is to choose the action (reject $H_0$ or not) that leads to the smaller posterior expected loss \citep{degroot}. Such test procedure rejects $H_0$ when
\begin{equation} \label{procedimento_test}
\tilde p_0=P(H_0 \textrm{ true } | \textbf{y}) \leq \frac{w_1}{w_0 + w_1}.
\end{equation}

It is common to use the Bayes factors (BF) for comparing a point hypothesis to a continuous alternative. We can define Bayes factor as the ratio of the posterior probabilities of the alternative and the null hypotheses over the ratio of the prior probabilities of the alternative and the null hypotheses. Thus, the BF is given by
\begin{equation} \label{BayesFactor}
BF = \frac{1-\tilde p_0}{\tilde p_0} \Bigg/ \frac{1-p_0}{p_0}.
\end{equation}

Observe that (\ref{BayesFactor}) would be the posterior odds against $H_0$ if $p_0=0.5$. Considering this situation, we can reconstruct the interpretation table of the BF given in \cite{Kass95} based on information of the posterior probability of separability $\tilde p_0$ and losses $w_0$ and $w_1$. Table \ref{tab:BF_test} presents the interpretation of the Bayesian test for separability proposed in this subsection.
\begin{table}
\begin{center}
\begin{small}
\begin{tabular}{cccc}
\hline
\multirow{2}{*}{BF} & \multirow{2}{*}{$\tilde p_0$} & \multirow{2}{*}{$(w_0; w_1)$} & Evidence against $H_0$ \\
 & & & (against separability) \\
\hline
1 to 3		& 0.50 to 0.25	& (1 to 3; 1)		& Not worth more than a bare mention \\
3 to 20	& 0.25 to 0.05	& (3 to 20; 1)	& Subtancial nonseparability \\
20 to 150	& 0.05 to 0.01	& (20 to 150; 1)	& Strong nonseparability \\
$>$ 150	& $<$ 0.01		& ($>$150; 1)	& Very strong nonseparability \\
\hline
\end{tabular}
\end{small}
\end{center}
\caption{Interpretation table for the Bayesian separability test. BF: Bayes Factor; $\tilde p_0$: posterior probability of separability; $w_0$: loss associated the decision of rejecting $H_0$ when $H_0$ is true; $w_1$: loss associated the decision of not rejecting $H_0$ when $H_1$ is true.}
\label{tab:BF_test}
\end{table}

Detailed information about BF is described in \cite{Kass95}. More details about Bayesian hypotheses testing can be seen in \cite{CRobert94} and \cite{Schervish95}. 

\subsection{Illustrative example}

We simulate four different scenarios from separable to very nonseparable structures. In this context, we consider the covariance model proposed in equation (\ref{eq:func_prop_repa}) with $\alpha_1 = \alpha_2 = 1$ and generate datasets with $p=2$ components, $n=80$ spatial locations and $T=20$ independent replicates in time. We consider a different degree of separability $\tilde \rho$ for each dataset and the same parameter specification $\boldsymbol{\Theta}=(\boldsymbol{\beta}_{1},\boldsymbol{\beta}_{2},\delta_{12}, \phi, \sigma_1, \sigma_2)$ with $\boldsymbol{\beta}_{1}=(1,-0.2,-0.8,0.5)$, $\boldsymbol{\beta}_{2}=(1.5,0.6,-0.5,-0.8)$, $\delta_{12}=1.5$, $b_{ij}=b_{ji}=\phi=0.05$, for $i,j=1,\ldots, p$ and $\sigma_1=\sigma_2=1$. We consider a Gaussian process, so $\textbf{y}_{t} \sim N_{np}(\boldsymbol{X\beta},\boldsymbol{\Sigma})$, $t = 1,\ldots, T$, where $\boldsymbol{\Sigma}$ is $np \times np$ covariance matrix and $\boldsymbol{X}$ are independent variables (latitude, longitude and altitude). The covariance function used is shown as follows in equation (\ref{eq:nsep_p2}).
\begin{equation}\label{eq:nsep_p2}
C_{ij}(\textbf{s},\xi)=\sigma_{i} \sigma_{j} \left(1+\delta_{ij}+\frac{h}{\phi}\right)^{-\alpha_{0}} \left(1+\frac{h}{\phi}\right)^{-1} \left(1+\delta_{ij}\right)^{-1} 
\end{equation}

\begin{figure}[htb]
\begin{center}
\subfigure[ref1][$\tilde \rho=0$]{\includegraphics[height=5cm]{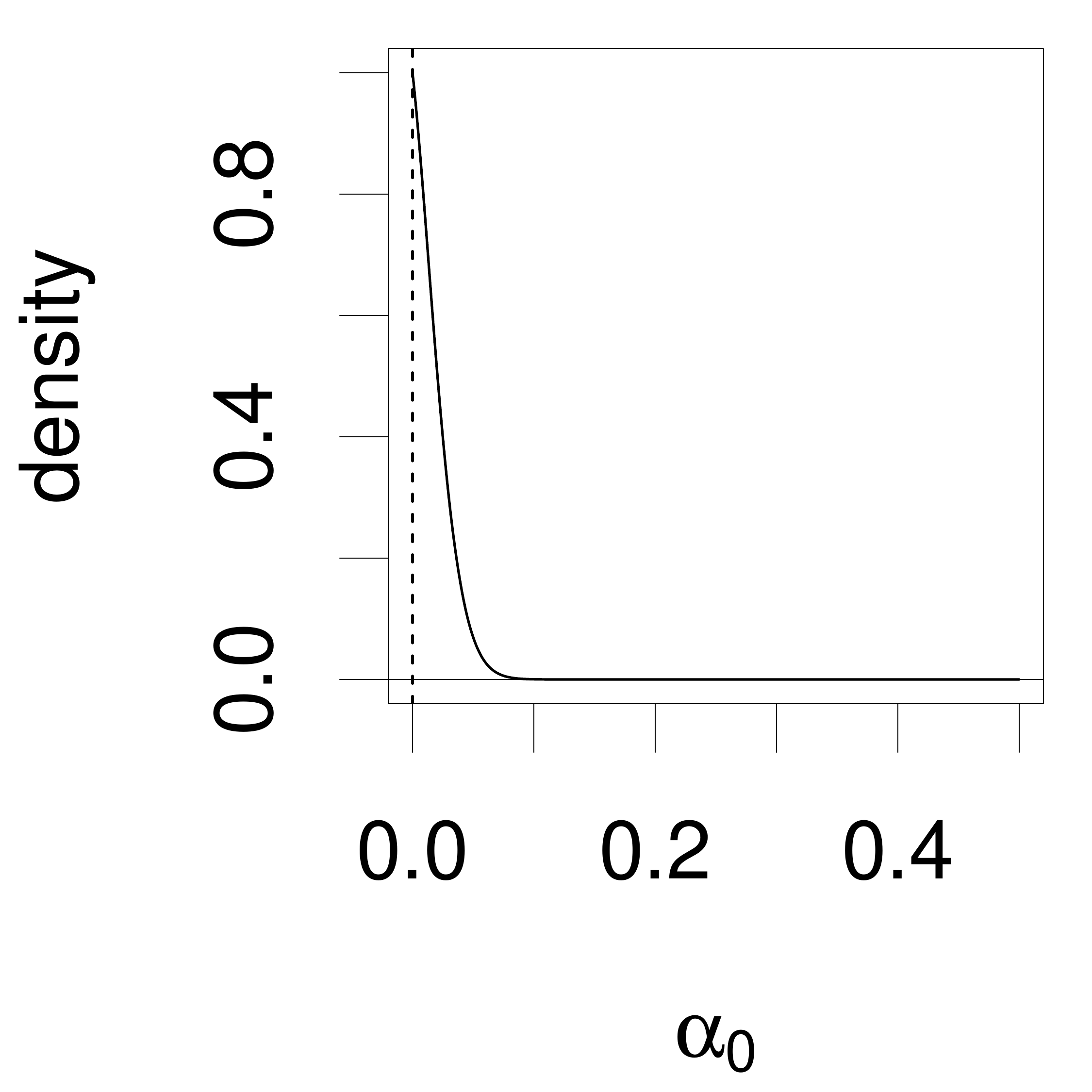}\label{GraphAp4-a}}
\subfigure[ref1][$\tilde \rho=0.05$]{\includegraphics[height=5cm]{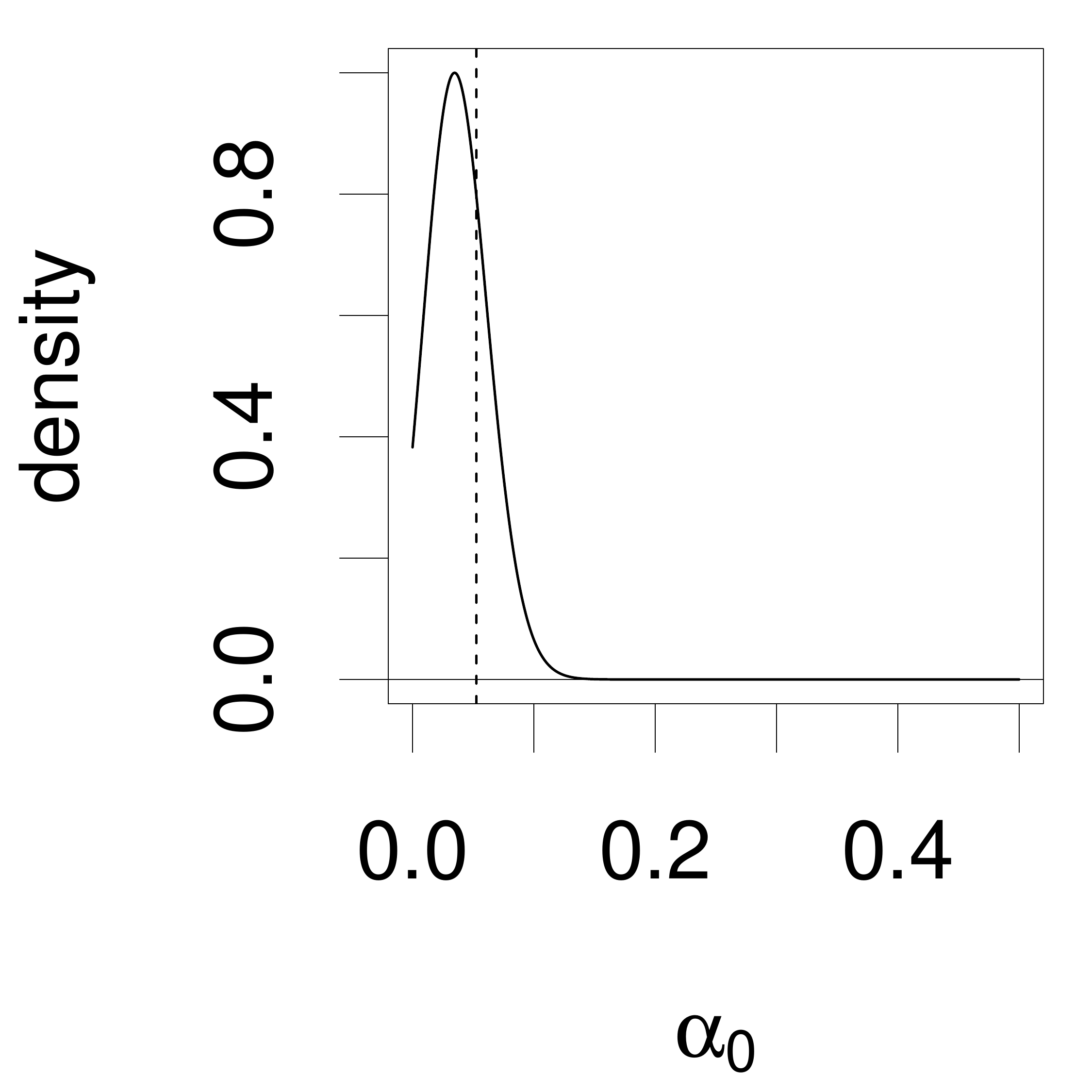}\label{GraphAp4-b}}\\
\subfigure[ref1][$\tilde \rho=0.10$]{\includegraphics[height=5cm]{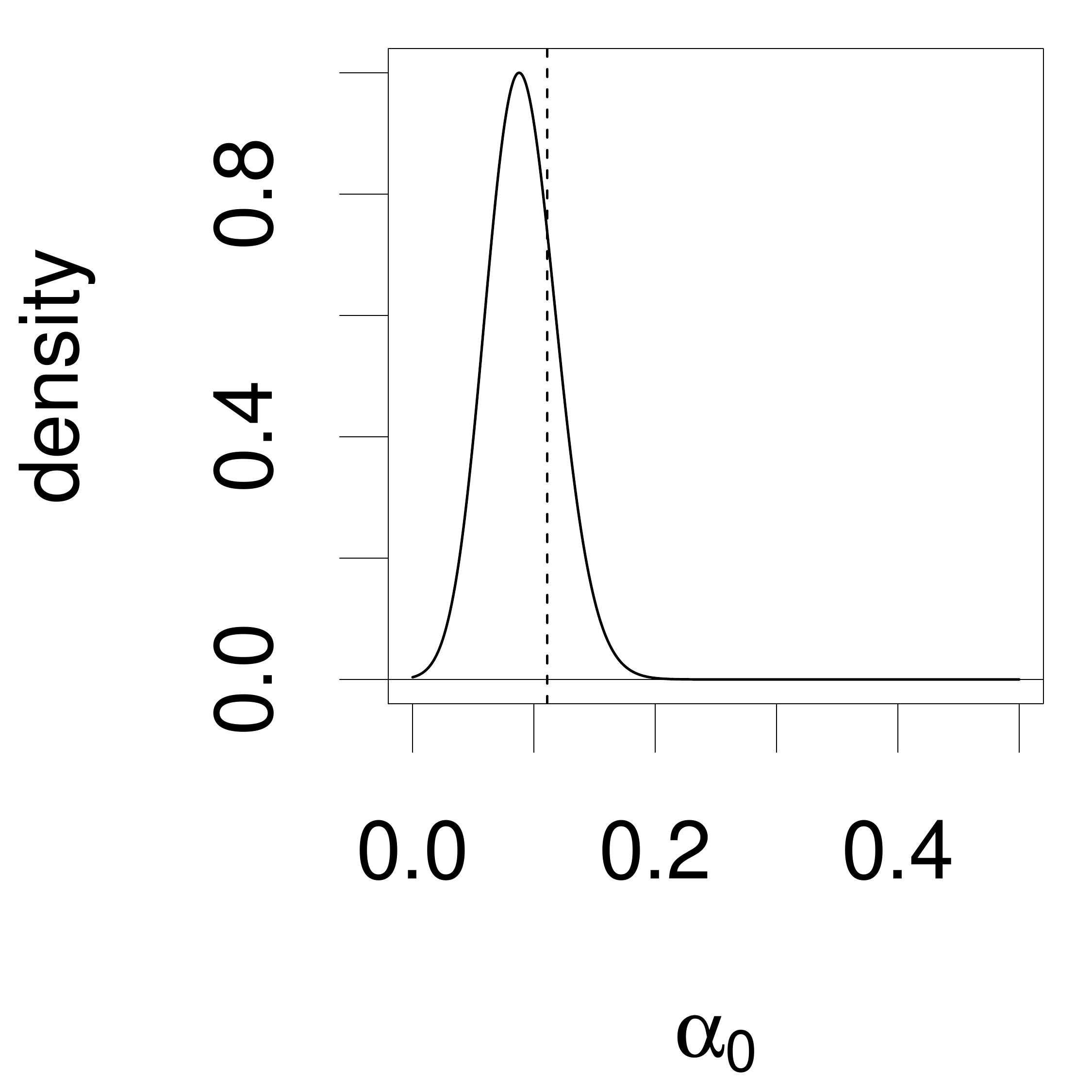}\label{GraphAp4-c}}
\subfigure[ref1][$\tilde \rho=0.20$]{\includegraphics[height=5cm]{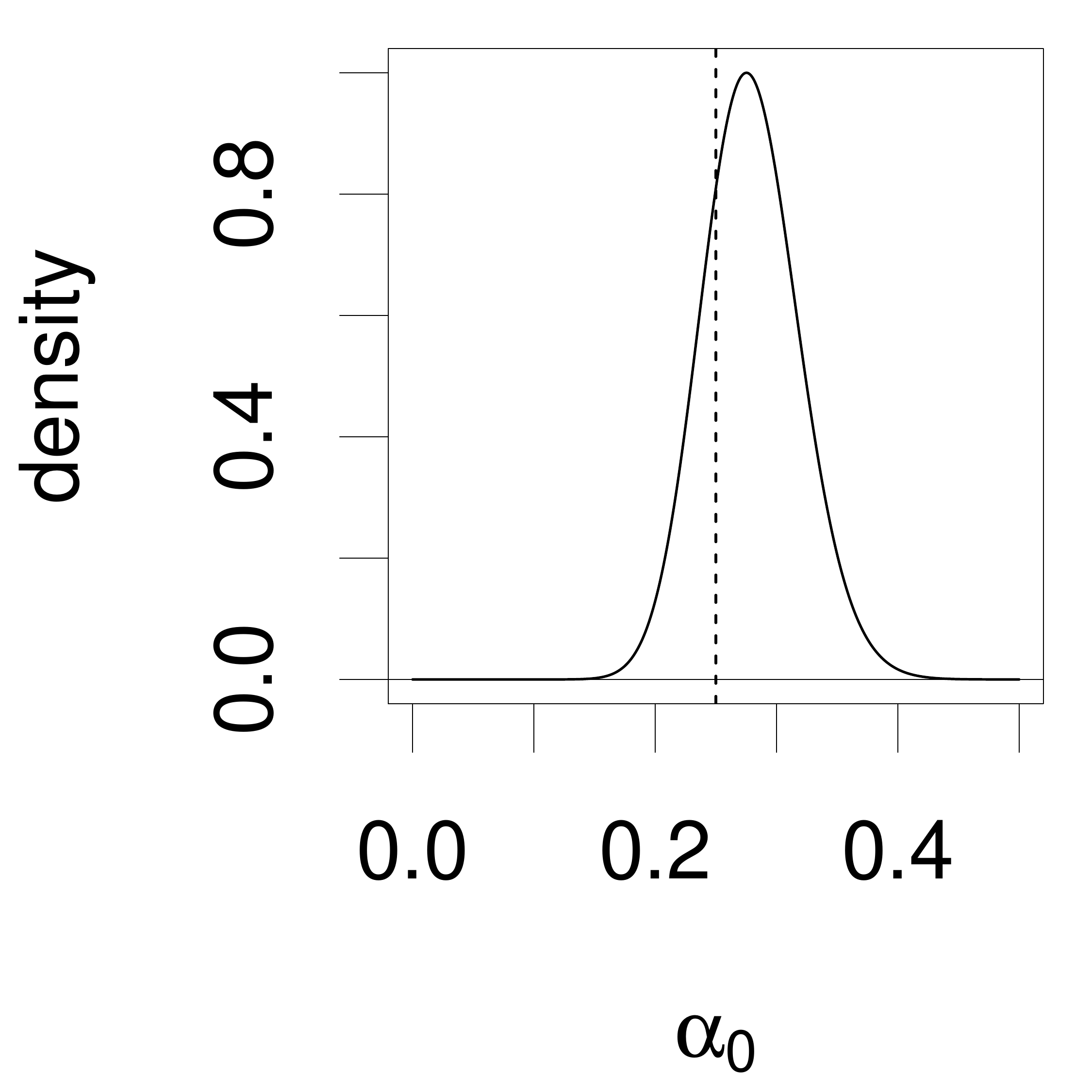}\label{GraphAp4-d}} 
\end{center}
\caption{Likelihood function described in (\ref{eq:verossimilhanca}) with covariance function defined in (\ref{eq:nsep_p2}). Dashed line: true value of separability parameter $\alpha_0$.} \label{GraphAp4}
\end{figure}

Figure \ref{GraphAp4} shows the likelihood function for $\alpha_{0}$ based on the degree of separability $\tilde \rho$. Note in Figure \ref{GraphAp4} that the data gives information regarding the estimation of separability. Furthermore, we expect the probability of separability to be very small when we define a dataset with $\tilde \rho=0.20$. Indeed in the fourth scenario, Figure \ref{GraphAp4-d}, the data indicates probability close to zero for the null hypothesis of separability, as presented in Tabel \ref{tab:BF_test}.

Table \ref{tab:prob_sep} presents the posterior probabilities $\tilde p_0$ for each model.  It is possible to see that the difference between the values of the measure of separability is subtle but the difference between the posterior probabilities is substantial. Thus, the posterior probability of separability is a much easier measure to interpret, regarding inference on separability. Also notice that $\tilde \rho$ values greater than 0.20 indicate strong nonseparability.
\begin{table}[htb]
\begin{center}
\begin{small}					
\begin{tabular}{ccccc}
\hline
$\tilde \rho$ & 0 & 0.05 & 0.10 & 0.20\\
\hline
$\tilde p_0$ & 0.987 & 0.854 & 0.251 & 0.035\\
\hline
\end{tabular}
\end{small}
\end{center}
\caption{Posterior probabilities of separability $\tilde p_0$ for each measure of separability $\tilde \rho$.}
\label{tab:prob_sep}
\end{table}

\subsection{Separability and Correlation}

In this section we show that high correlation between components implies separability.
Return to the scenario defined in equation (\ref{eq:cov_y_sep}) where we have a variable \textbf{X} with spatial correlation matrix \textbf{R} and a variable \textbf{Y}, conditioned on variable \textbf{X}, with spatial correlation structure defined by \textbf{S} matrix. Therefore, using the results of equation (\ref{eq:cov_y_sep}) and considering the transformation (\ref{eq:validando}) to evaluate the correlation structure of \textbf{Y} we have 
\begin{equation}\label{eq:corr_y_sep}
Corr[Y_{i}, Y_{j}]  = \textbf{S}_{ij} - (a_{12}^{*})^{2}\textbf{S}_{ij} + (a_{12}^{*})^{2}\textbf{R}_{ij}
\end{equation}
with $a_{12}^{*}=\frac{a_{12}}{[a_{11}a_{22}]^{1/2}}$.

From equation (\ref{eq:corr_y_sep}), note that the separable structure is obtained when $\textbf{S}=\textbf{R}$, as presented in Section \ref{section2}, or when the correlation between the components, $a_{12}^{*}$, is 1 or -1. Therefore, high dependence between variables implies separability, in other words, indicates proportionality between spatial structures. If we work with highly correlated variables, the distances $\delta$'s will be estimated in values close to zero since we are assuming a strong similarity between the components. Consequently, the posterior probability of separability $\tilde p_0$ will be close to one, indicating separable structure.

\section{Simulated examples} \label{simulation}

This section presents a simulated example for two different scenarios to verify the relation between our separability parameter and posterior probabilities of separability.

We use the covariance function defined in (\ref{eq:nsep_p2}) and generate two datasets with $p=3$ components, $n=55$ spatial locations in the $[0,1] \times [0,1]$ square and $T=20$ independent replicates in time. The information about three spatial locations were removed for prediction. The datasets are given by\
\vspace{0.3cm}

\noindent \textbf{Dataset 1}: We consider the separable specification of the function in (\ref{eq:nsep_p2}), i.e, we define $\tilde \rho = 0$, that implies $\alpha_0=0$. The $\delta$ parameters were chosen such that the variables present high correlation. Thus, we consider the following parameter specification $\boldsymbol{\Theta}=(\boldsymbol{\beta}_{1},\boldsymbol{\beta}_{2},\boldsymbol{\beta}_{3},\sigma_1,\sigma_2,\sigma_3,\delta_{12},\delta_{13},\delta_{23},\phi)$ with $\boldsymbol{\beta}_{1}=(1,-0.2,-0.8,0.5)$, $\boldsymbol{\beta}_{2}=(1.5,0.6,-0.5,-0.8)$, $\boldsymbol{\beta}_{3}=(1.8,-0.4,-0.3,0.6)$, $\sigma_1=2$, $\sigma_2=-1$, $\sigma_3=2.5$, $\delta_{12}=0.1$, $\delta_{13}=0.2$, $\delta_{23}=0.15$ and $\phi=0.1$. \
\vspace{0.3cm}

\noindent \textbf{Dataset 2}: We define $\tilde \rho = 0.18$, that implies $\alpha_0 \approx 0.22$.  The $\delta$ parameters were chosen such that the variables present weak/moderate correlation. The $\boldsymbol{\beta}=(\boldsymbol{\beta}_{1},\boldsymbol{\beta}_{2},\boldsymbol{\beta}_{3})$ parameters are the same defined in dataset 1. Thus, we consider the following parameter specification $\sigma_1=2$, $\sigma_2=1$, $\sigma_3=2.5$, $\delta_{12}=2$, $\delta_{13}=2.2$, $\delta_{23}=1.9$ and $\phi=0.1$\
\vspace{0.3cm}

In order to illustrate the correlations between the variables in each dataset, we estimate the parameters in the model $\textbf{y} = \boldsymbol{\mu}+\boldsymbol{\epsilon}$ for each spatial location $i$, $i=1,\ldots, 55$, that is,
\begin{equation}
\textbf{y} = (\textbf{y}_{1},\textbf{y}_{2},\textbf{y}_{3})' \sim N_3(\boldsymbol{\mu},\textbf{A}), \hspace{0.5cm}.
\end{equation}

\begin{figure}[htb]
\begin{center}
\subfigure[ref1][Dataset 1]{\includegraphics[height=5.4cm]{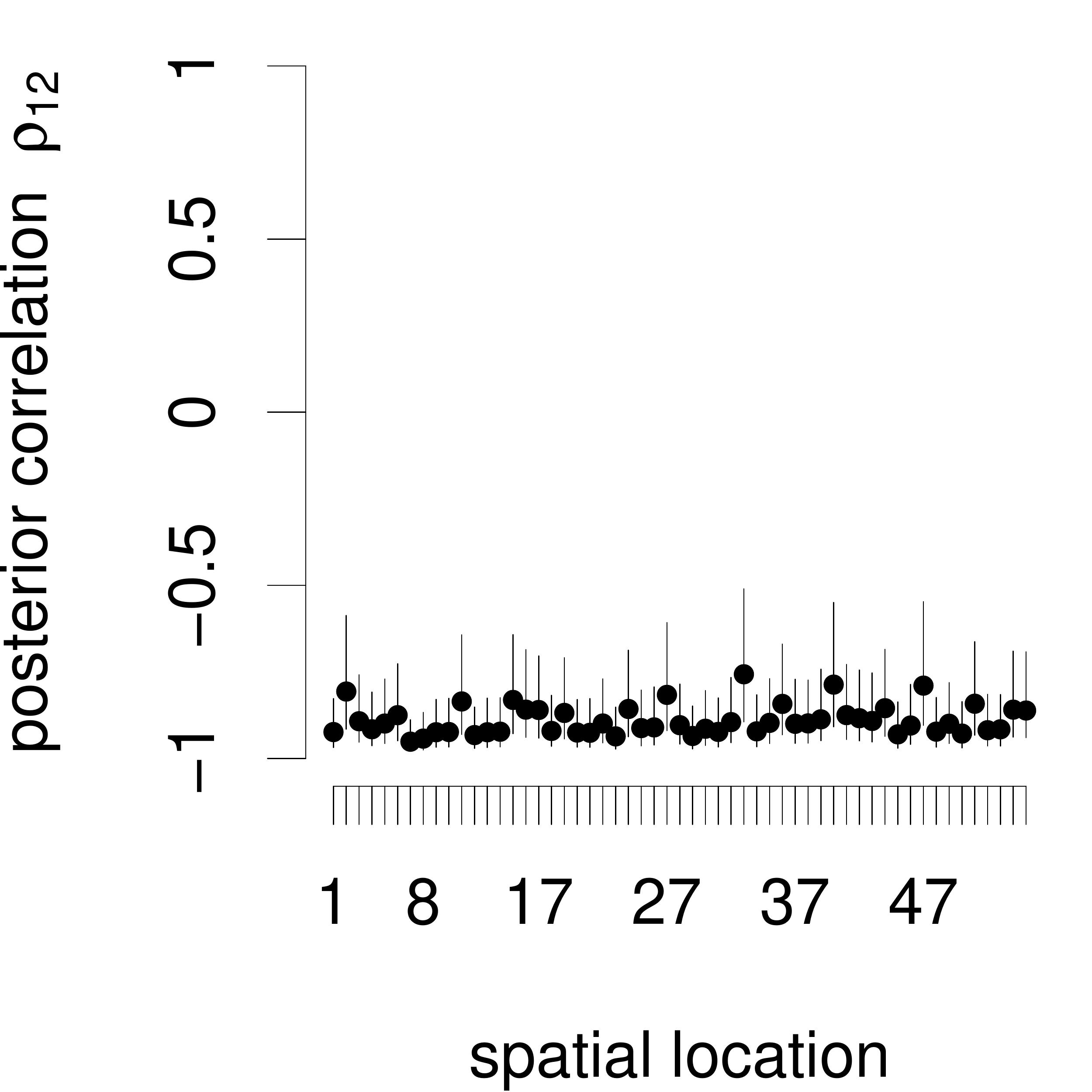}\label{GraphAp61-a}
			              \includegraphics[height=5.4cm]{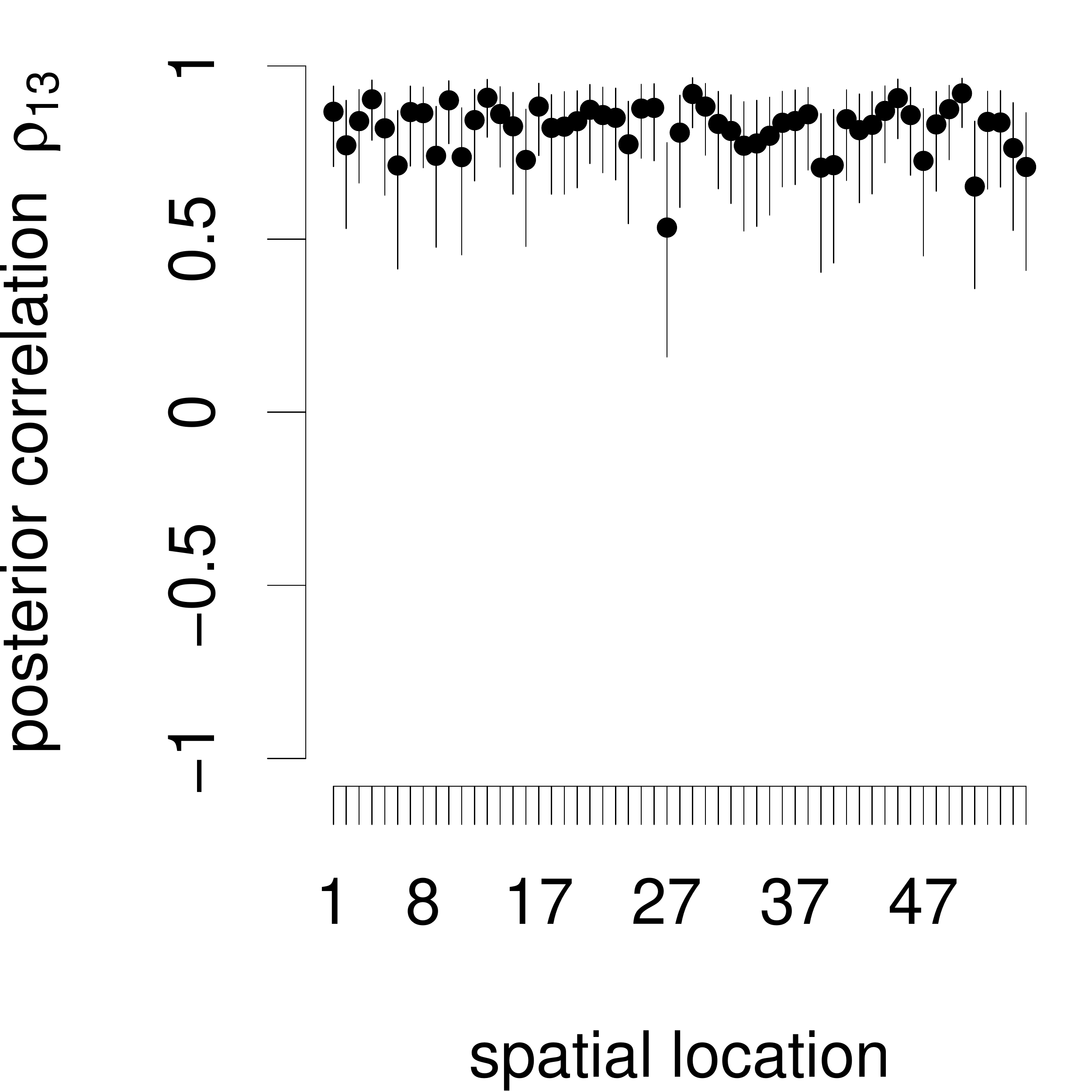}\label{GraphAp61-b}
       				   \includegraphics[height=5.4cm]{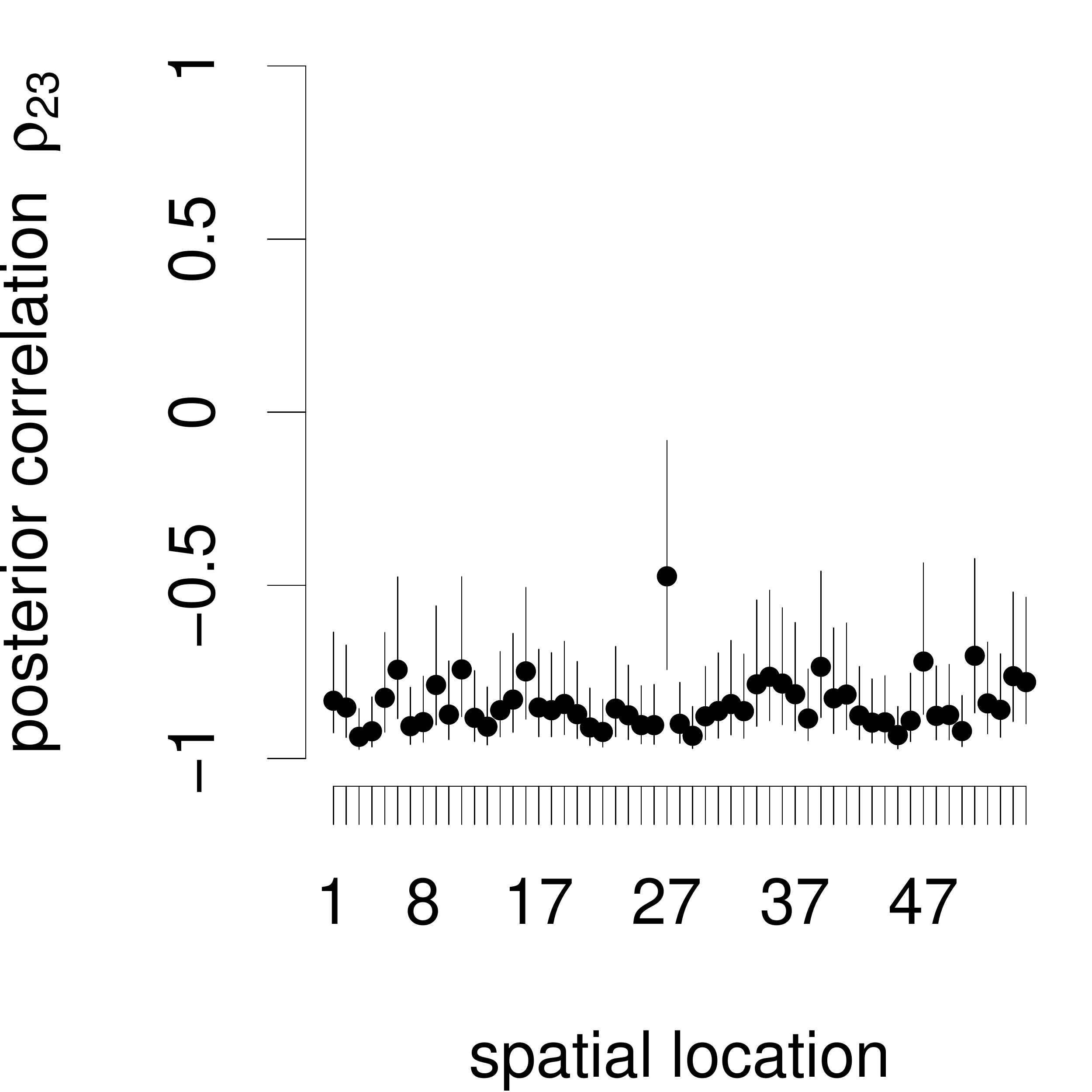}\label{GraphAp61-c}}

\subfigure[ref1][Dataset 2]{\includegraphics[height=5.4cm]{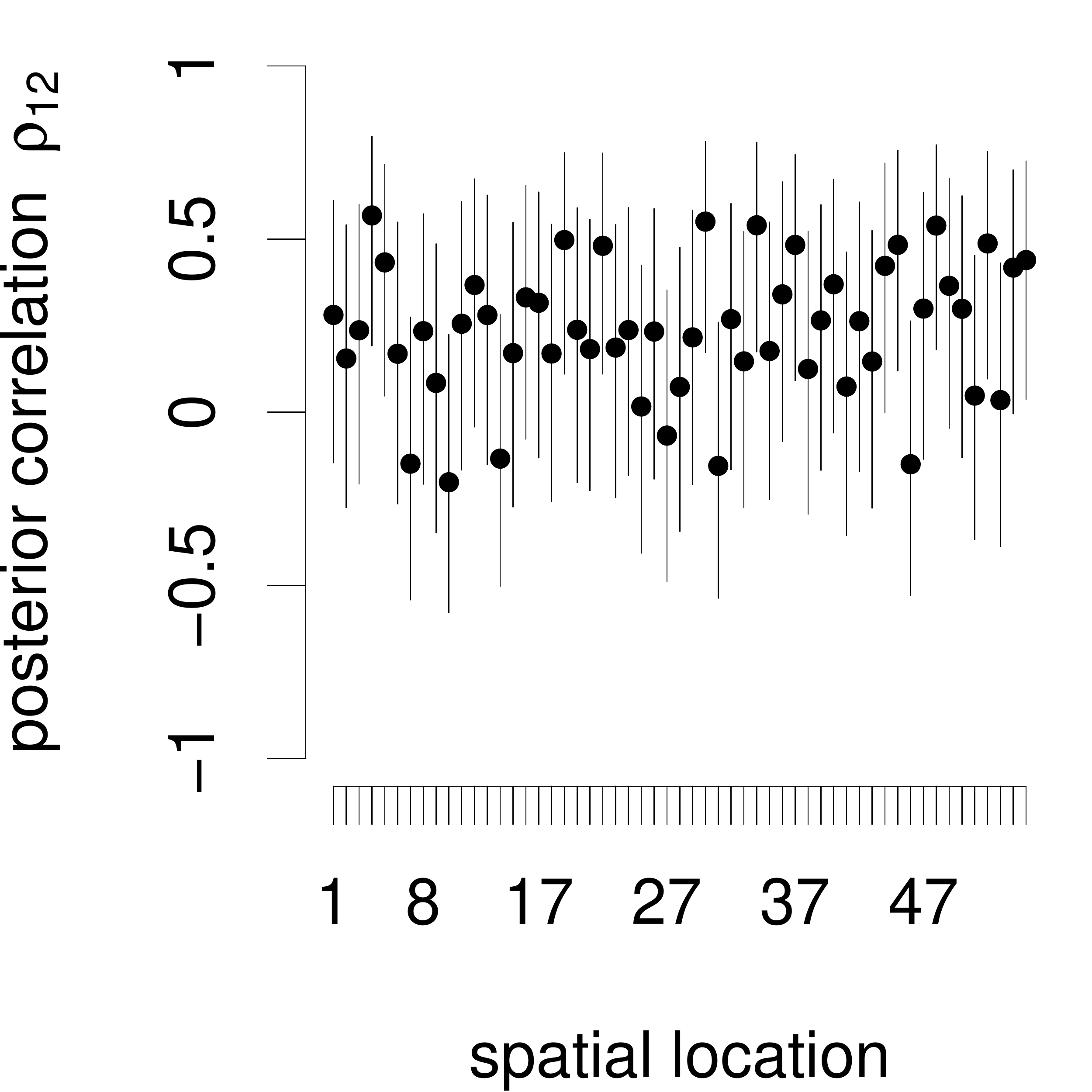}\label{GraphAp62-a}
				   \includegraphics[height=5.4cm]{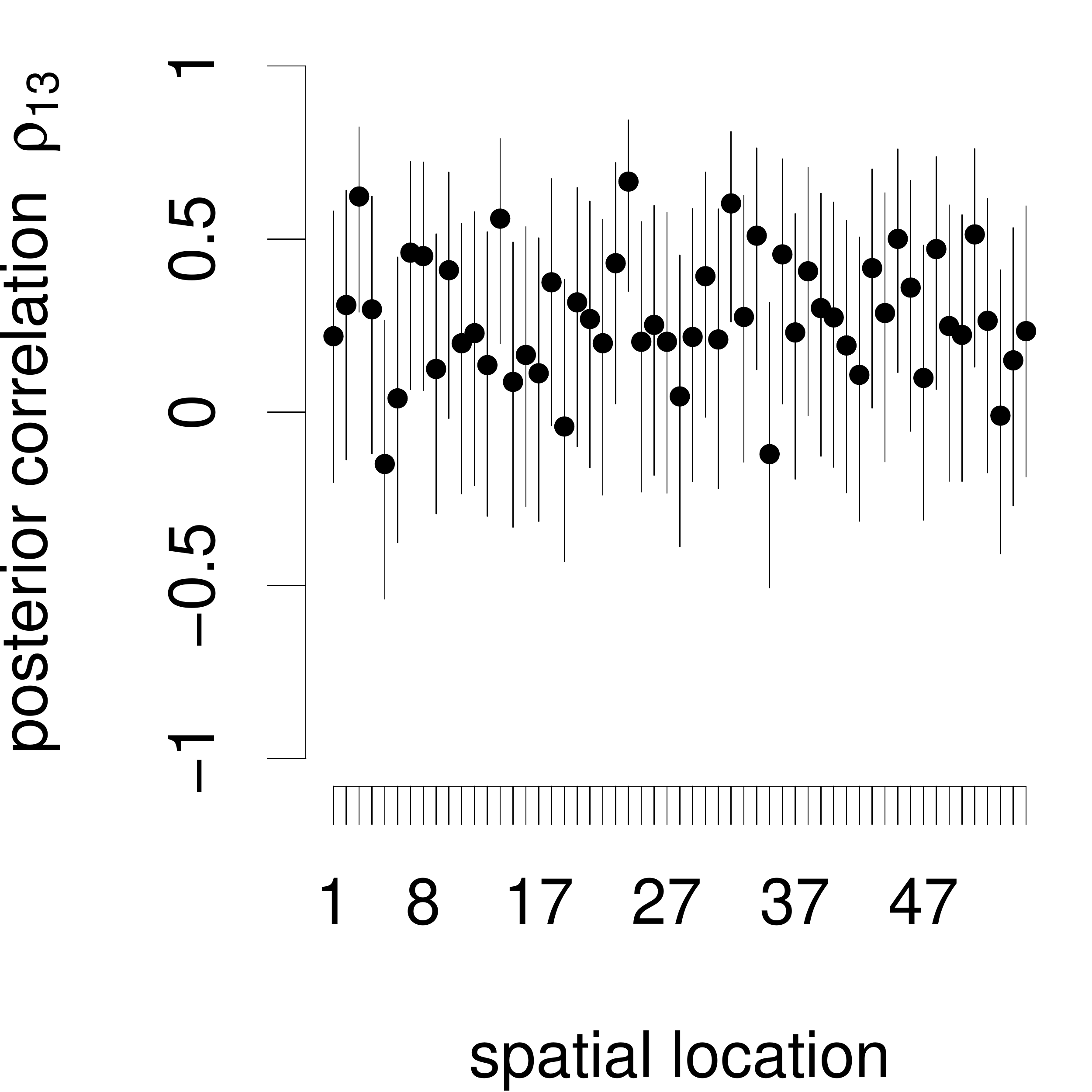}\label{GraphAp62-b}
       				   \includegraphics[height=5.4cm]{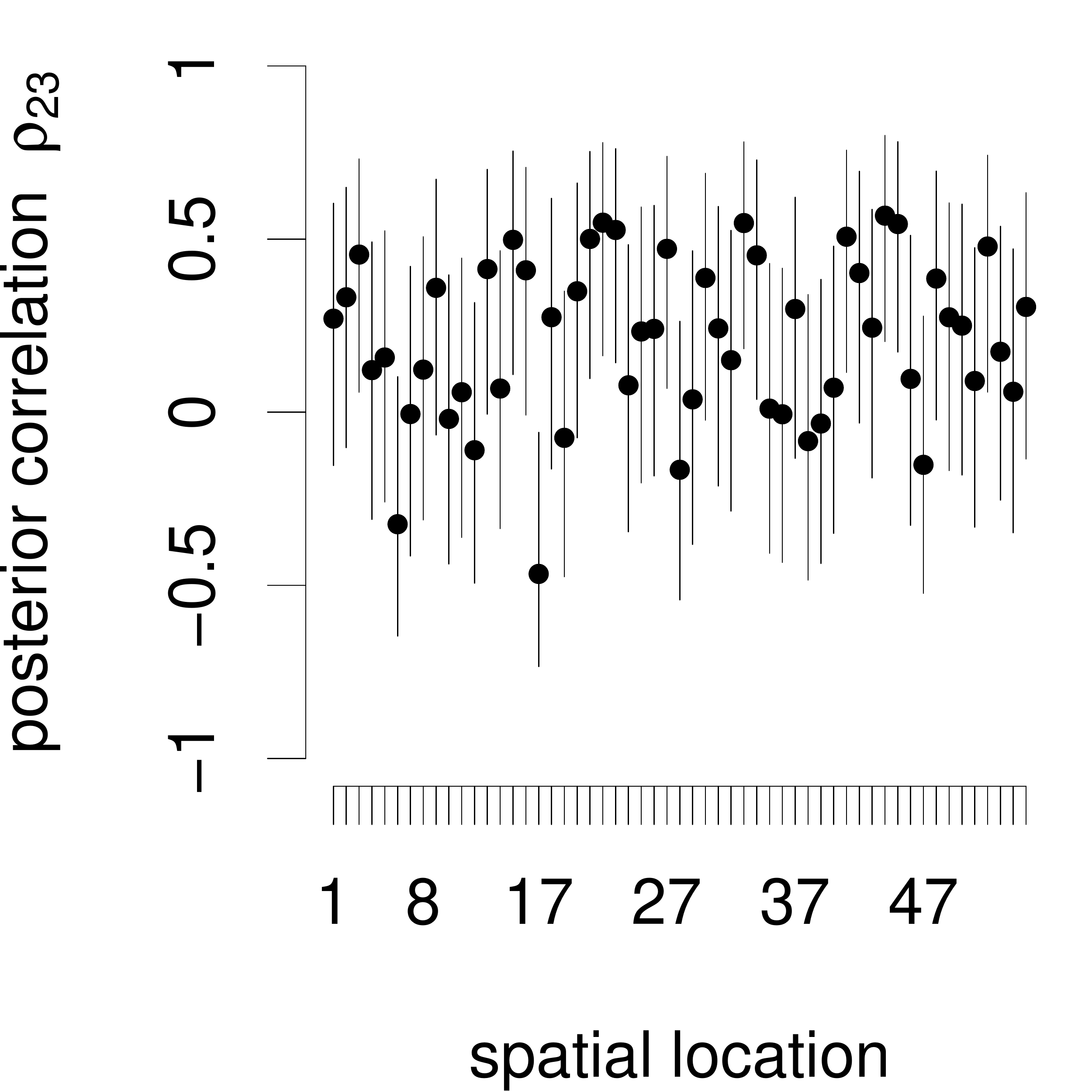}\label{GraphAp62-c}}
\end{center}
\caption{Posterior median and 95\% CI of the cross-correlations among variables for each dataset.} \label{GraphAp61}
\end{figure}

Figure \ref{GraphAp61} presents the posterior median and $95\%$ credibility interval of the correlations between variables for each dataset.

We estimate three multivariate models for each dataset and their performances are compared in predictive terms. We consider the following models: the separable model with covariance function in the Cauchy family, as presented in \ref{Covs}; the nonseparable model with covariance function defined in (\ref{eq:nsep_p2}) and a continuos prior for $\alpha_0$; and the nonseparable model with covariance function defined in (\ref{eq:nsep_p2}) and a mixture prior for $\alpha_0$, given by a point mass at zero and a continuos function for $\alpha_0>0$.

The priors for the parameters in the proposed model for all datasets follow the discussion in Section \ref{sec_inference}. We assume that $\sigma_{i} \sim N(0,100)$, $i = 1,\ldots, 3$, $\delta_{ij} \sim Ga(1,0.5)$,  $i \neq j$, $i,j = 1,\dots, 3$, $\phi \sim Ga(0.75\times med(d_{s}),0.75)$, with $med(d_{s})=0.5145$, $\boldsymbol{\beta} \sim N_{12}(\textbf{0},1000\textbf{I}_{12})$. For the nonseparable model that consider a continuous prior for $\alpha_0$, assuming that $\alpha_0 \sim Ga(1,1)$ and for the second nonseparable model which considers a mixture prior for $\alpha_0$ we assume a point mass at zero and a $Ga(1,3)$ for $\alpha_0>0$ both with the same weight. For the separable model the priors are as follows: $\textbf{A}\sim InverseWishart(\textbf{I}_{3},4)$, $\phi \sim Ga(0.75\times med(d_{s}),0.75)$, with $med(d_{s})=0.5145$ and $\boldsymbol{\beta} \sim N_{12}(\textbf{0},1000\textbf{I}_{12})$. Inference was carried under a MCMC scheme and for convergence monitoring we use the algorithms present in the \texttt{Coda} package in the \texttt{R} \citep{Plummer06}.

\begin{figure}[htb]
\begin{center}
\includegraphics[height=5.4cm]{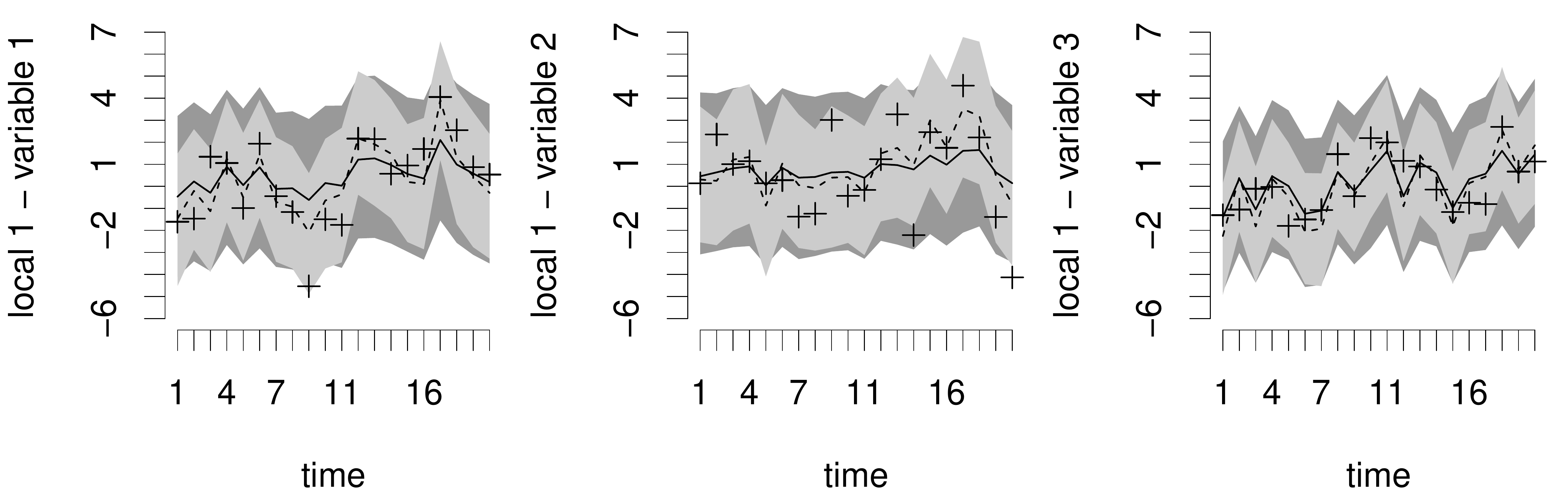}
\includegraphics[height=5.4cm]{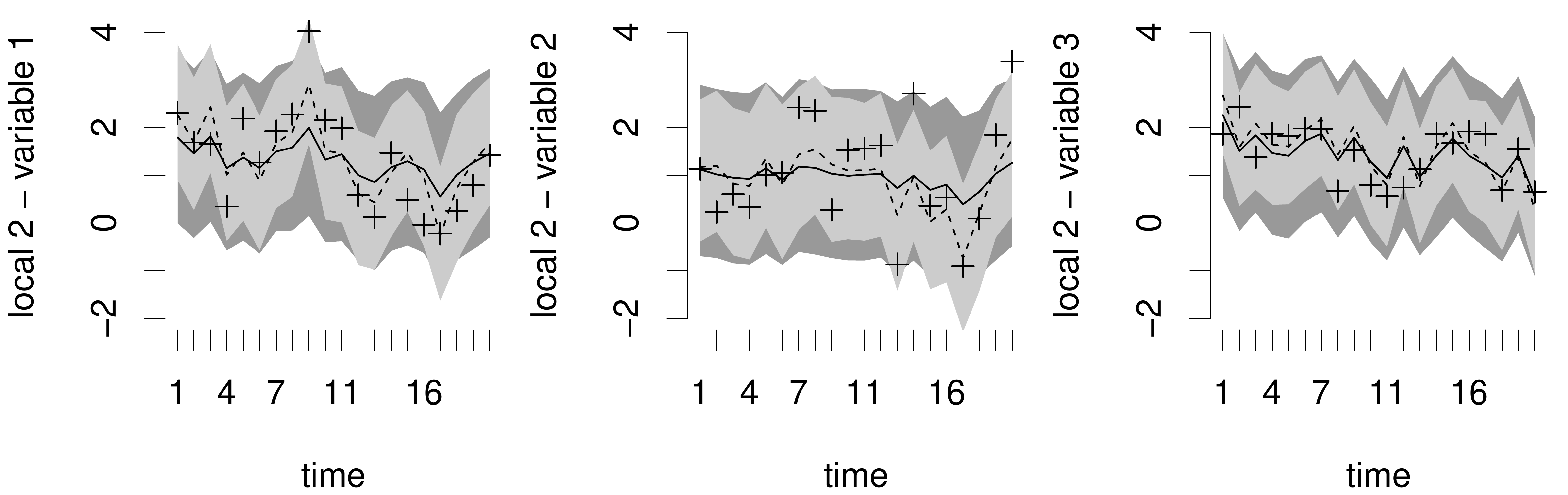}
\includegraphics[height=5.4cm]{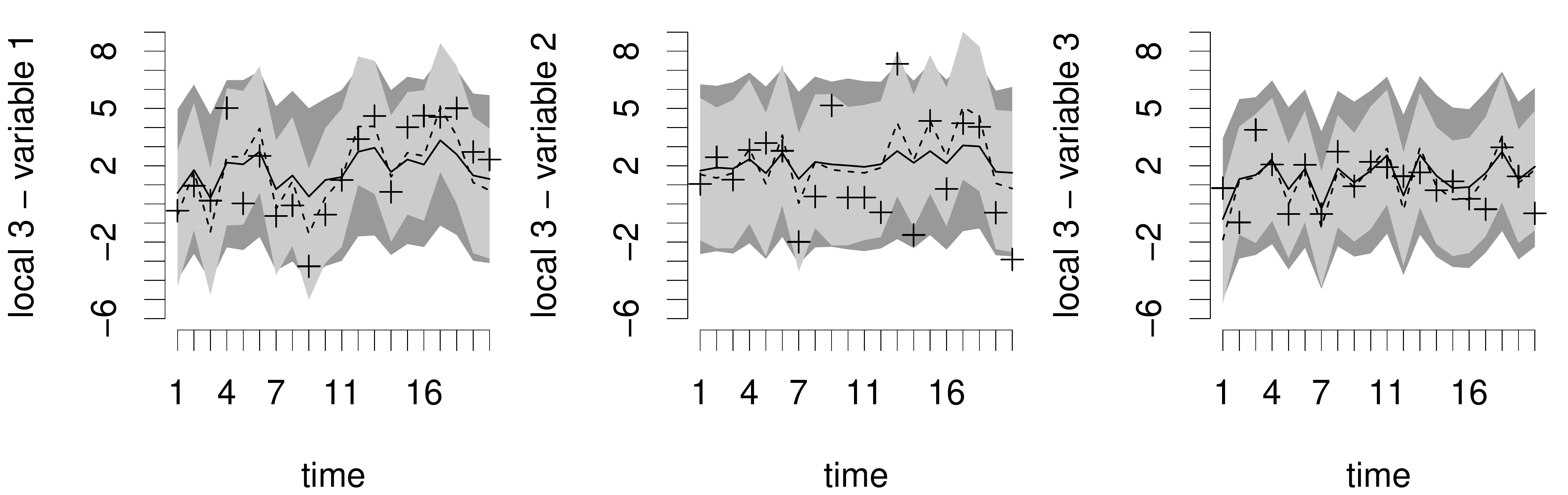}
\end{center}
\caption{Prediction of separable model and nonseparable model with mixture. Data: separable ($\tilde \rho=0$); points: true values; full line: prediction mean of the separable model; dashed line: prediction mean of the nonseparable model; dark gray area: 95\% CI of the separable model; light gray area: 95\% CI of the nonseparable model. } \label{GraphAp6_pred-rho000}
\end{figure}

\begin{figure}[htb]
\begin{center}
\includegraphics[height=5.4cm]{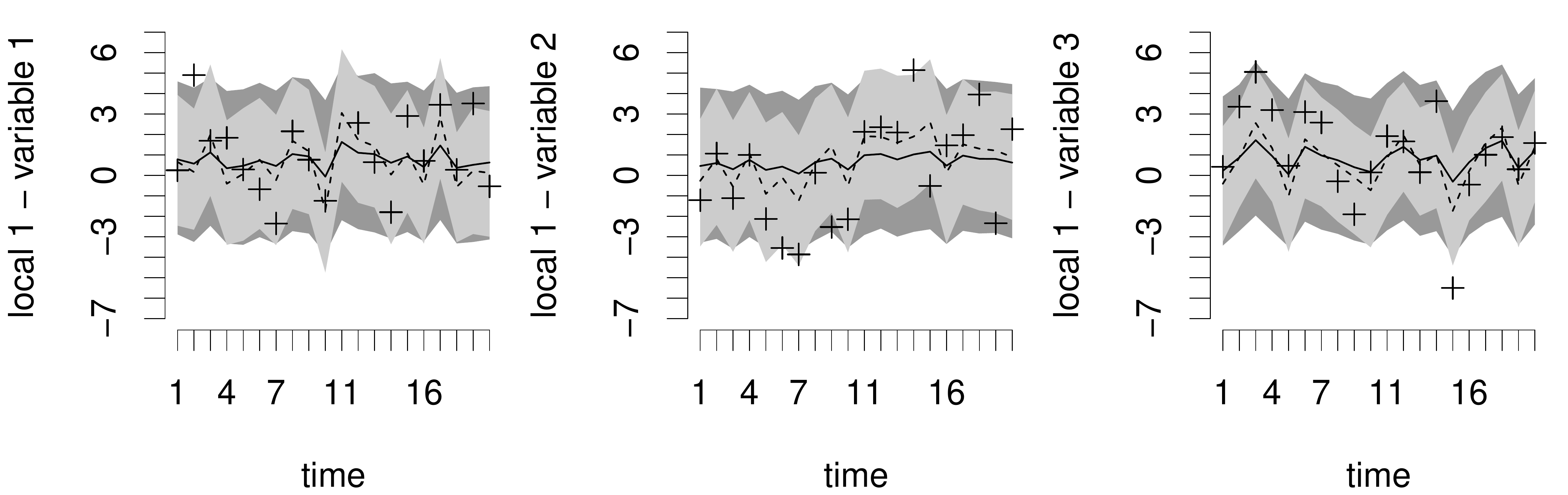}
\includegraphics[height=5.4cm]{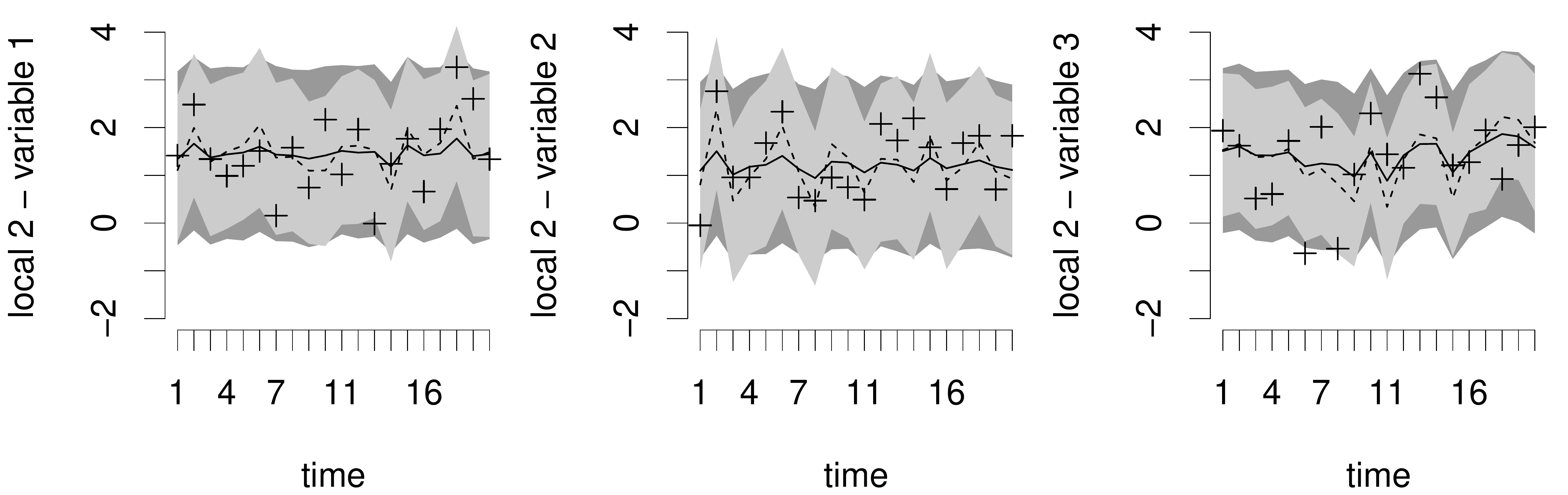}
\includegraphics[height=5.4cm]{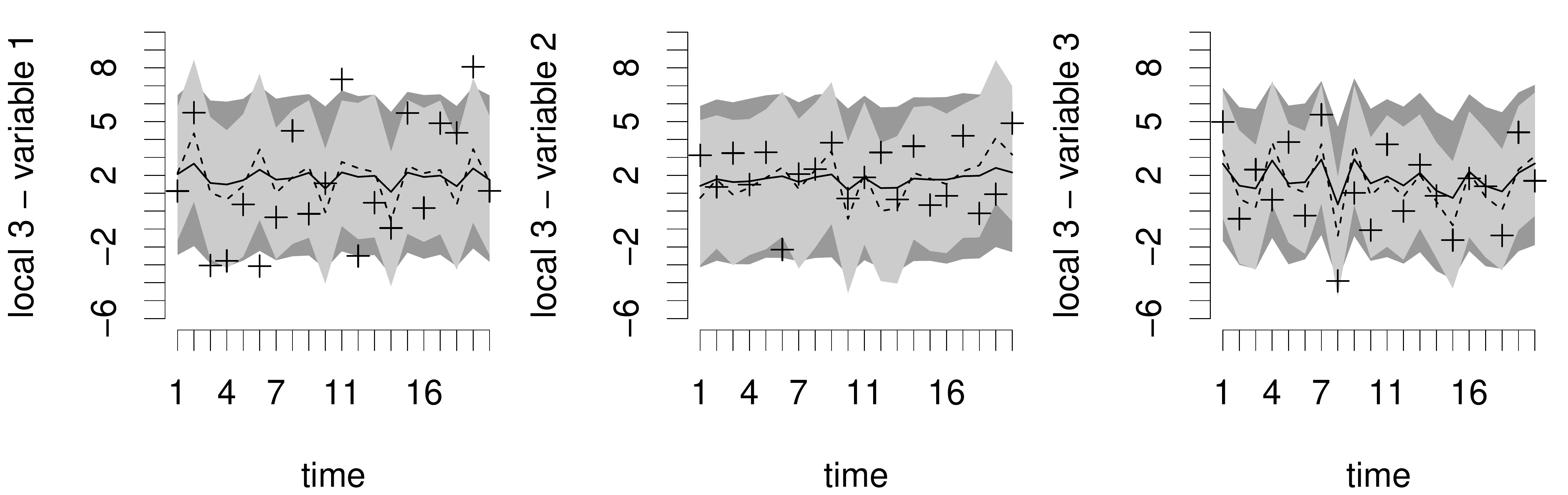}
\end{center}
\caption{Prediction of separable model and nonseparable model with mixture. Data: nonseparable ($\tilde \rho=0.18$); points: true values; full line: prediction mean of the separable model; dashed line: prediction mean of the nonseparable model; dark gray area: 95\% CI of the separable model; light gray area: 95\% CI of the nonseparable model.} \label{GraphAp6_pred-rho018}
\end{figure}

Table \ref{Tab_simula} presents predictive measures for model comparison for each dataset. The IS (Interval Score) and LPML (Logarithm of the Pseudo Marginal Likelihood) comparison measures are described in \cite{GneitRaf07} and \cite{Ibrahim01b}, respectively. These measures are detailed in \ref{Comp}. Note that the separable model always presents the worst predictive performance. Figures \ref{GraphAp6_pred-rho000} and \ref{GraphAp6_pred-rho018} present, for each dataset, the mean predictions and their respective 95\% credibility intervals, for the separable and nonseparable (with mixture) models. For all datasets, the predictions of the nonseparable model with mixture seem to present point estimates closer to the true values. In addition, the uncertainty associated with the prediction of the nonseparable model with mixture is always smaller than that of the separable model. Thus, we note that even for data with separable structure, the nonseparable model with mixture seems to be the best option.
\begin{table}[H]
\begin{center}
\begin{small}	
\begin{tabular}{ccccc}
\hline
Data & Model & average IS & LPML & $\tilde p_0$\\
\hline
\multirow{2}{*}{Separable}   		& Separable 			     	  	& 250.30 & -6,171.81 & --\\
\multirow{2}{*}{($\tilde \rho = 0$)}	& Nonseparable (without mixture)	& 197.93 & -4,834.59 & --\\
         				   		& Nonseparable (with mixture)	  	& 195.58 & -4,764.78 & 0.927\\
\hline
\multirow{2}{*}{Nonseparable}   	& Separable 			     	  	& 261.88 & -7,859.68 & --\\
\multirow{2}{*}{($\tilde \rho = 0.18$)}	& Nonseparable (without mixture) 	& 218.15 & -7,103.68 & --\\
         				   		& Nonseparable (with mixture)	  	& 217.92 & -6,939.44 & 0.163\\
\hline
\end{tabular}
\end{small}
\end{center}
\caption{Predictive measures for model comparison and posterior probability of separability for the simulation examples.}
\label{Tab_simula}
\end{table}

\section{Cear\'a weather dataset} \label{application}

\begin{figure}[htb]
\begin{center} 
\includegraphics*[height=6cm,width=6cm]{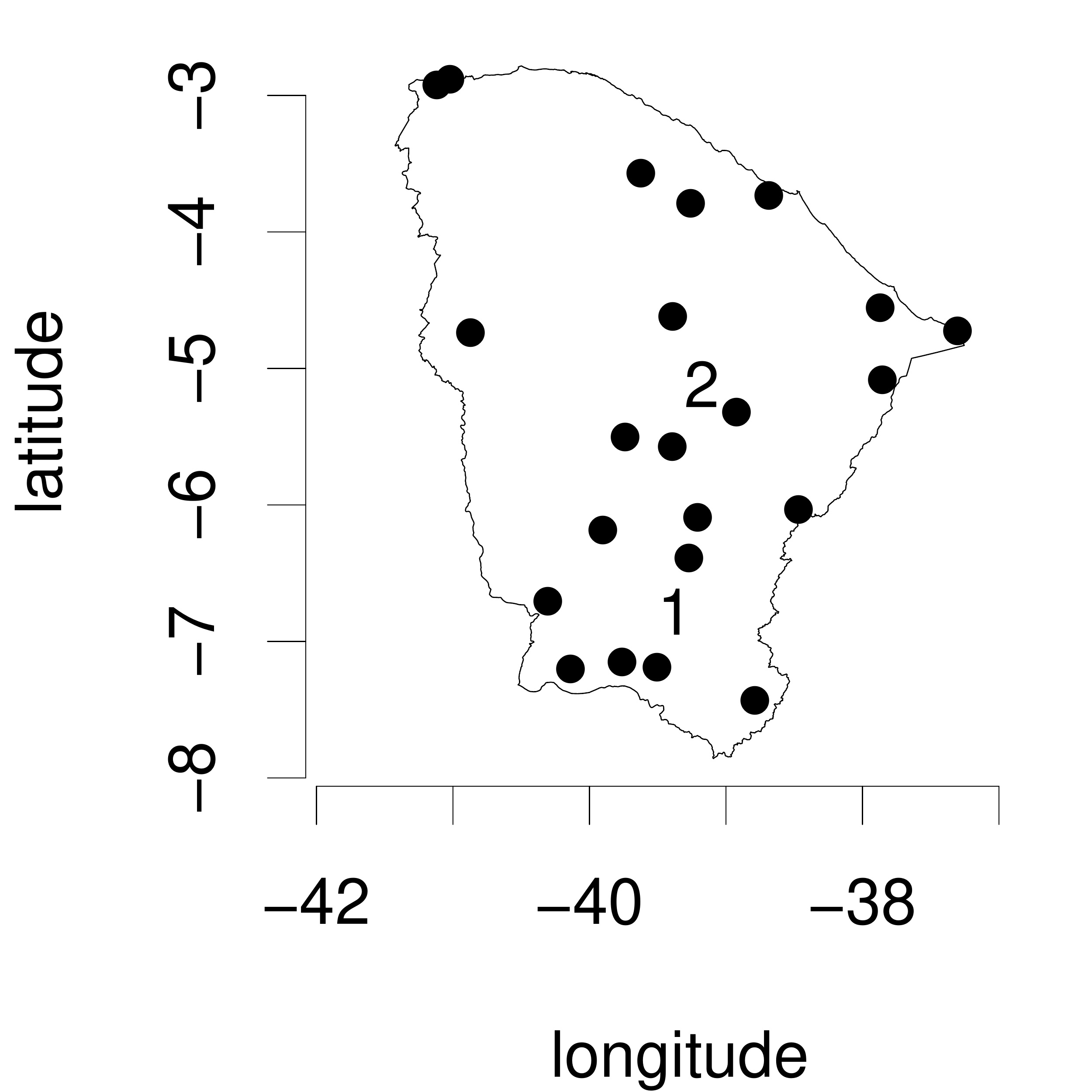} 
\caption{Collection of monitoring stations in Cear\'a state, Brazil. Numbers 1 and 2 are spatial locations considered for the predictive comparison.}
\label{GraphAp5}
\end{center}
\end{figure}

In this section we apply the model defined in (\ref{eq:nsep_p2}) to an illustrative dataset obtained for a collection of monitoring stations in Cear\'a state, Brazil. The weather dataset were obtained from Instituto Nacional de Pesquisas Espaciais (INPE) and consists of three variables, \textit{temperature} ($^{\circ}$C), \textit{humidity} (\%) and \textit{solar radiation} (MJ/$m^{2}$), measured daily at 12 o'clock and recorded at 24 stations from December 20, 2010 to February 28, 2011. 
Locations with less than 10\% missings have gone through an imputation process\footnote{The imputation was performed applying the \texttt{mice} package in \texttt{R}.}. In addition, we work with the seasonally adjusted series to obtain $T=71$ independent replicates in time. For predictive comparison and validation, we consider two spatial locations.
Figure \ref{GraphAp5} shows the locations of these 24 monitoring sites and two hold-out sites on a latitute-longitude scale.

In order to evaluate the correlations between the variables, we estimate the parameters in the model $\textbf{y} = \boldsymbol{\mu}+\boldsymbol{\epsilon}$ for each spatial location $i$, $i=1,\ldots, 24$, that is,
\begin{equation}
\textbf{y} = (\textbf{y}_{temperature},\textbf{y}_{humidity},\textbf{y}_{radiation})' \sim N_3(\boldsymbol{\mu},\textbf{A}), \hspace{0.5cm}.
\end{equation}

\begin{figure}[htb]
\begin{center}
\subfigure[ref1][temp. $vs.$ humid.]{\includegraphics[height=5.4cm]{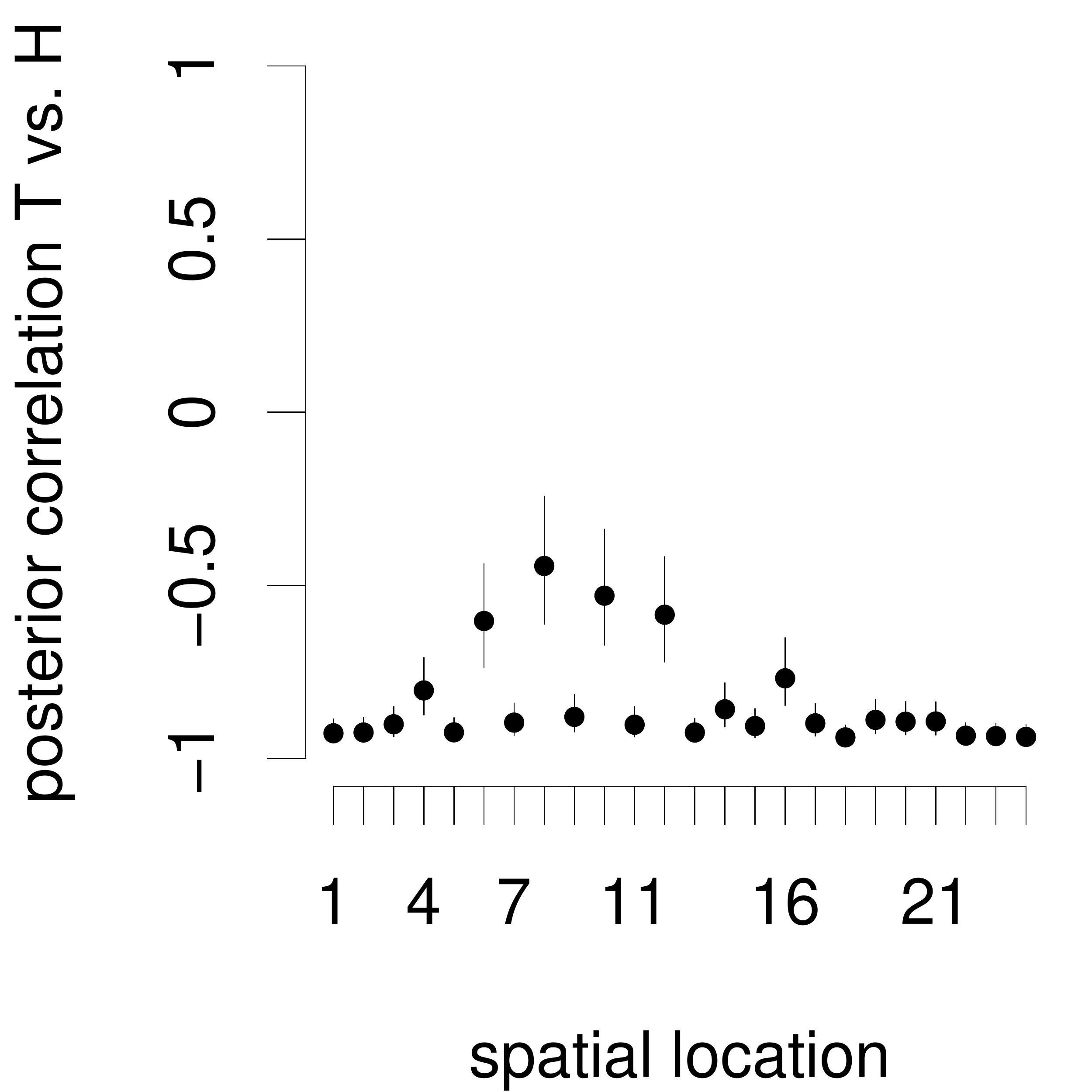}\label{GraphAp6A-a}}
\subfigure[ref1][temp. $vs.$ solar rad.]{\includegraphics[height=5.4cm]{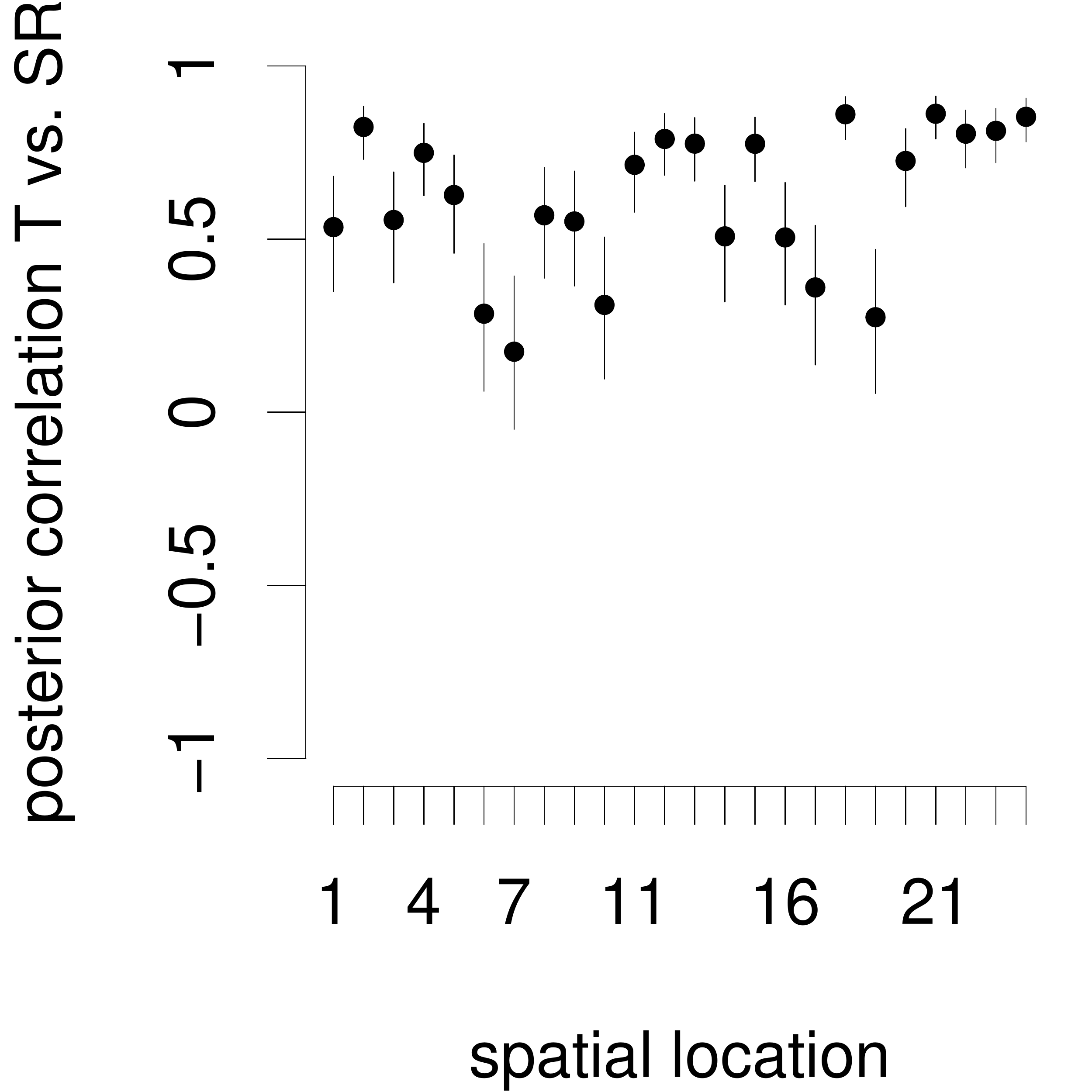}\label{GraphAp6A-b}}
\subfigure[ref1][humid. $vs.$ solar rad.]{\includegraphics[height=5.4cm]{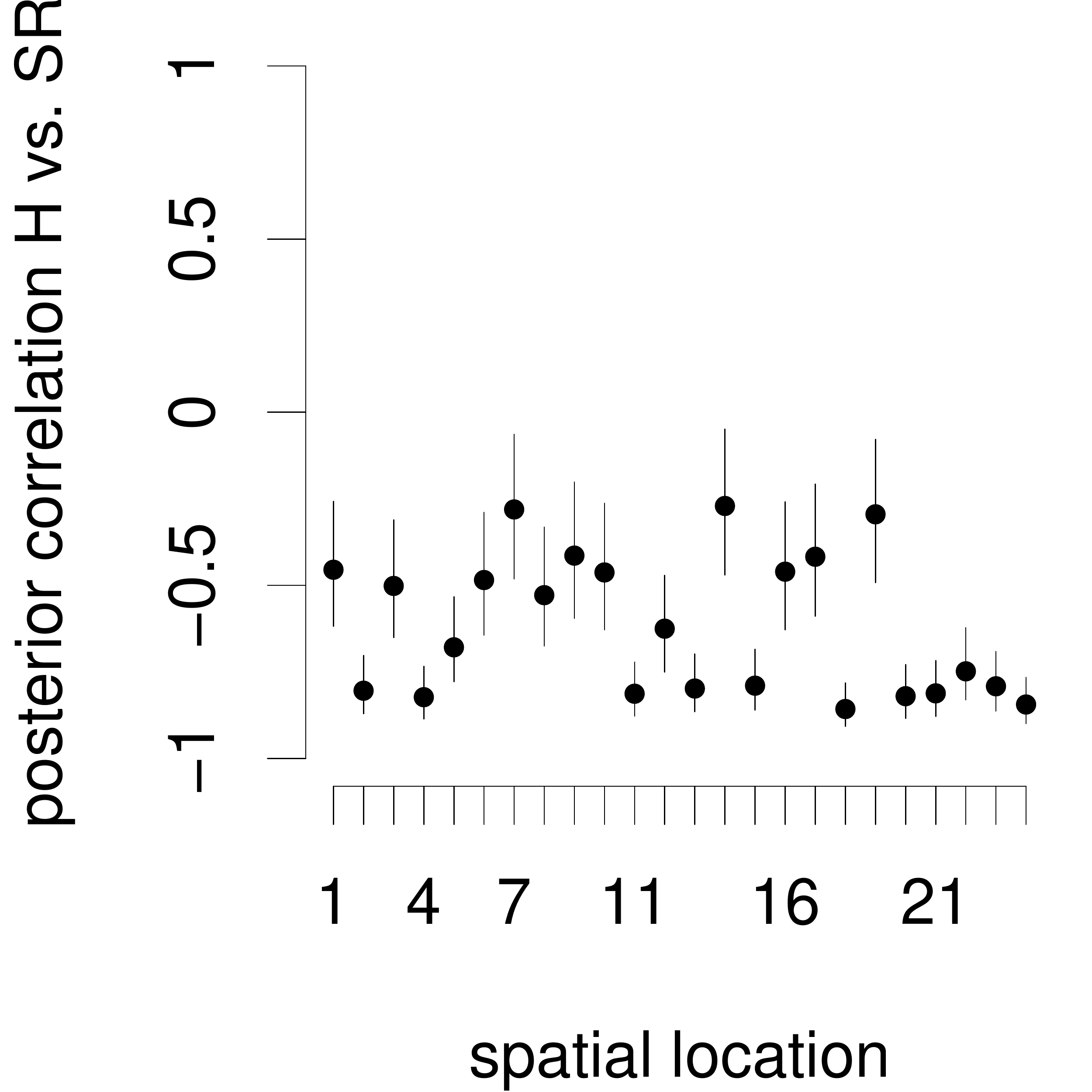}\label{GraphAp6A-c}} 
\end{center}
\caption{Posterior median and 95\% CI of the cross-correlations among components.} \label{GraphAp6A}
\end{figure}

From Figure \ref{GraphAp6A} note that there is strong correlation between the three variables. Indeed, the variables are very similar, and so we expect the component distance between them to be small.

We compare the multivariate model defined in (\ref{eq:nsep_p2}) with a separable multivariate model and the independent univariate models for each variable. The covariance functions used in the univariate and separable models belong to the Cauchy family and are presented in \ref{Covs}. Parameter estimation was performed considering the likelihood described in (\ref{eq:verossimilhanca}). 
The priors for the parameters in the proposed model follow the discussion in Section \ref{sec_inference}. So, we assume that $\sigma_{i} \sim N(0,100)$, $i = 1,\ldots, 3$, $\delta_{ij} \sim Ga(1,0.5)$,  $i \neq j$, $i,j = 1,\dots, 3$, $\phi \sim Ga(0.05\times med(d_{s}),0.05)$, with $med(d_{s})=1.958$, $\boldsymbol{\beta} \sim N_{12}(\textbf{0},1000\textbf{I}_{12})$. We consider two nonseparable models: the first sets a continuous prior for $\alpha_0$ ($\alpha_0 \sim Ga(1,0.25)$) and the second model sets a mixture prior for $\alpha_0$, given by a point mass at zero and a $Ga(1,3)$ for $\alpha_0>0$. For the separable model the priors are as follows: $\textbf{A}\sim InverseWishart(\textbf{I}_{3},4)$, $\phi \sim Ga(0.25\times med(d_{s}),0.25)$, with $med(d_{s})=1.958$ and $\boldsymbol{\beta} \sim N_{12}(\textbf{0},1000\textbf{I}_{12})$. For the univariate models the prior distributions are given by: $\theta \sim Ga(1,0.25)$, with $\theta=\frac{1}{\sigma^{2} }$, $\phi \sim Ga(0.1\times med(d_{s}),0.1)$, with $med(d_{s})=1.958$ and $\boldsymbol{\beta} \sim N_{4}(\textbf{0},1000\textbf{I}_{4})$. The simulation method used was the MCMC. For the convergence monitoring we use the algorithms present in the \texttt{Coda} package in the \texttt{R}.

Table \ref{tab:ERROR} presents predictive measures for model comparison. Note that working with the three variables without considering the dependence between them is not a good option. In predictive terms, the independent model has the worst performance. As presented in Figure \ref{GraphAp6A}, the variables present considerably large correlation. Therefore, it is expected that the proposed model presents a high probability of separability. Note that $\tilde p_0 = 1$ indicates complete separability, but to obtain this result the correlation between the variables must be 1 or -1. The estimated $\tilde p_0$ was 0.912, indicating small but non zero probability of a nonseparable structure. As the dependence between the variables is not perfect, the proposed nonseparable model must present better performance than the fully separable model, since it admits to work with both structures. Furthermore, if we do not consider the possibility of $\alpha_0$ being equal to zero, that is, if we work with a completely nonseparable model (without mixture), then the predictive performance will still be substantially better than the separable model as confirmed by the IS and LPML in Table \ref{tab:ERROR}. 
\begin{table}[htb]
\begin{center}
\begin{small}					
\begin{tabular}{lrr}
\hline
Model & average IS & LPML\\
\hline
Independent & 1,851.56 & -15,311.36\\
Separable & 658.85 & -14,377.12 \\
Nonseparable (without mixture) & 610.57 & -14,165.68\\
Nonseparable (with mixture) & 608.06 & -13,980.60\\
\hline
\end{tabular}
\end{small}
\end{center}
\caption{Comparison of models in predictive terms for the Cear\'a weather dataset.}
\label{tab:ERROR}
\end{table}

\begin{figure}[htb]
\begin{center}
\subfigure[ref1][local 1]{\includegraphics[height=14cm]{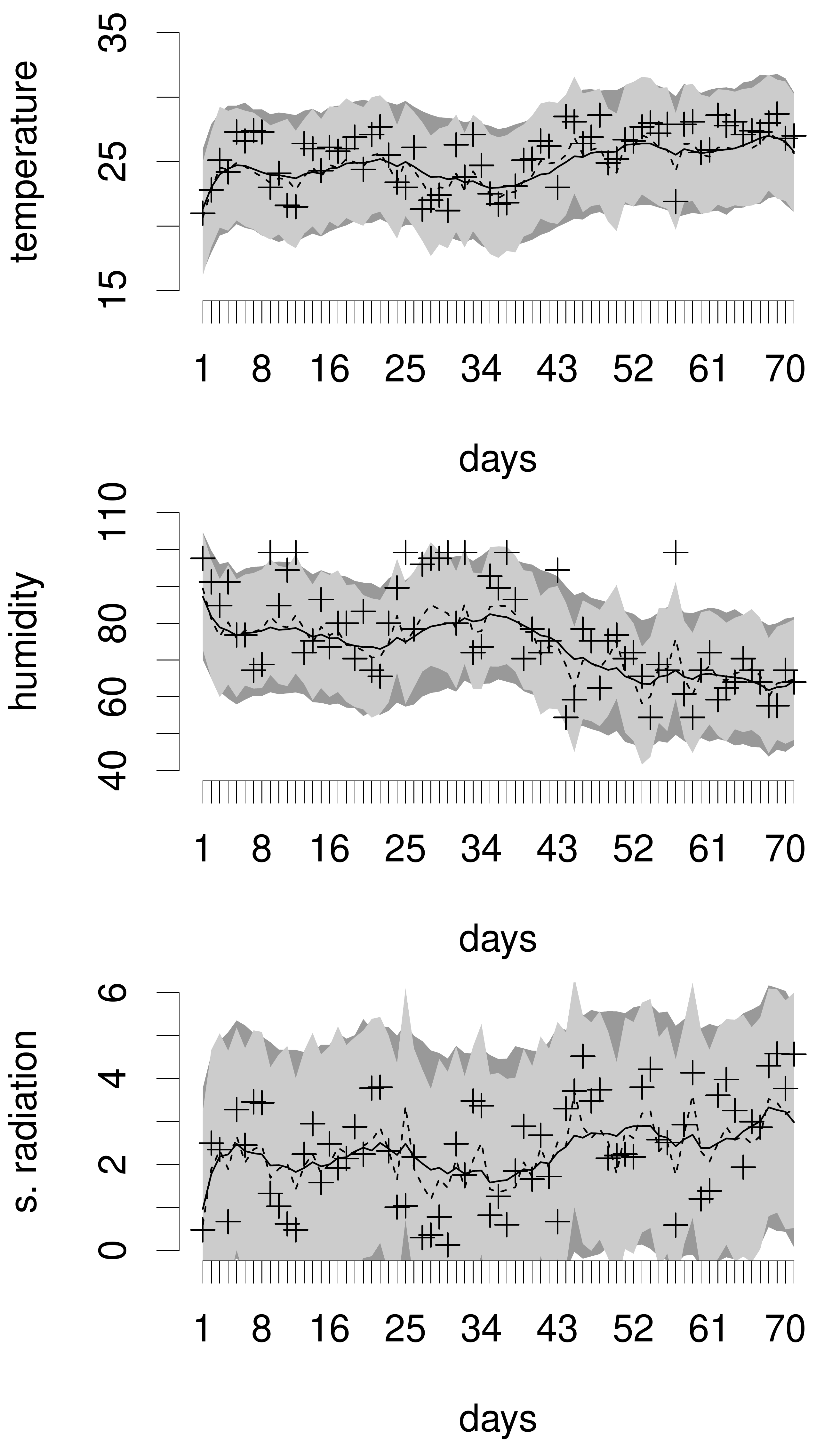}}
\subfigure[ref1][local 2]{\includegraphics[height=14cm]{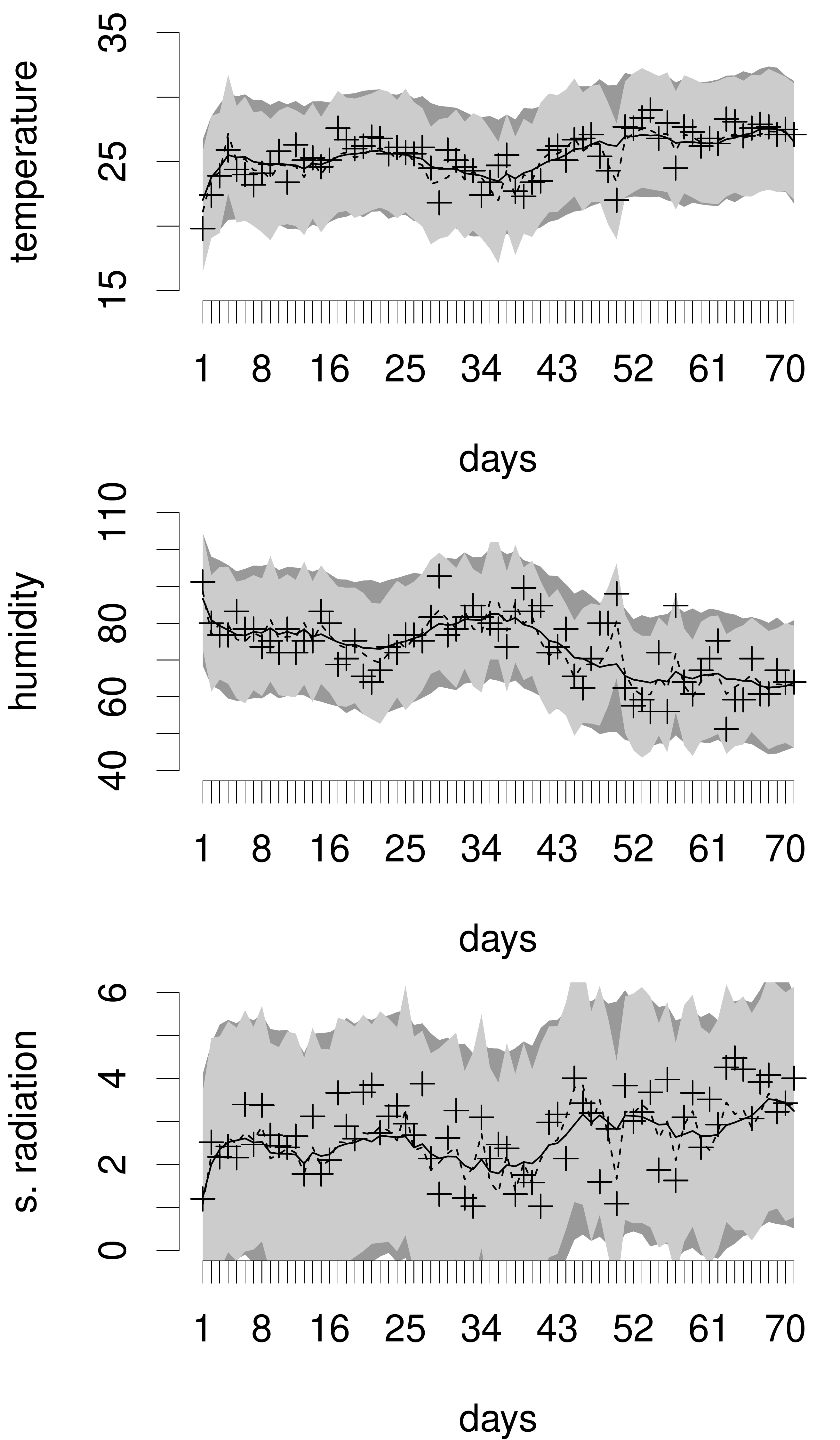}}
\end{center}
\caption{Prediction of separable model and nonseparable model with mixture. Points: true values; full line: prediction mean of the separable model; dashed line: prediction mean of the nonseparable model; dark gray area: 95\% CI of the separable model; light gray area: 95\% CI of the nonseparable model.} \label{GraphAp7}
\end{figure}

The proposed model that considers a weighting between separability and nonseparability presented better predictive results. Figure \ref{GraphAp7} presents the predicted mean and the 95\% CI for the temperature, humidity and solar radiation in the separable model and for the nonseparable model with mixture. Note that the predictive fit across all variables is better in the nonseparable model. In addition, this model presents lower uncertainty for the forecasts.

\section{Discussion} \label{conclusions}

We have proposed a new flexible class of covariance functions for multivariate spatial processes based on the convex combination of separable covariance functions and on latent dimensions. 

The proposed class is defined through a bivariate random vector, regardless of the dimension of the vector of observed components. Furthermore, it is derived from a valid covariance function which is able to achieve small to large degrees of nonseparability.

We have proposed a Bayesian test to measure the degree of separability between space and components. From the posterior probabilities $\tilde p_0$ we can choose the most suitable model. Indeed, the proposed measure is easier to interpret than the separability parameter itself. We verified that high correlation between variables implies separability. The proposed model by \cite{Gent10}, which uses the function presented in \cite{Gneit02}, does not present realistic results. In their illustration, the variables present moderate/high correlation and the parameter that measures the degree of separability indicates strong nonseparability. That result is contradictory. Indeed the separability parameter in their covariance function, when at the upper limit, does not imply high nonseparability. For more details see \cite{FonsSteel17}.

In the illustration, we verified that the nonseparable model with a mixture prior for $\alpha_0$ presents the better predictive results and lower uncertainty for the forecast, since it considers a weighting between the separable and nonseparable models. Even without considering the mixture prior, the nonseparable models present better performance than the separable one.

An important discussion must be made about latent distances $\delta$'s. These parameters measure the dissimilarity between the variables. Indeed, if we have variables that are highly correlated, that is, that are similar to each other, we expect the distance between them to be close to zero. Another possible approach is to consider the latent vectors $\xi$'s instead of latent distances in the model. Note that estimating these vectors might reduce the number of parameters to be estimated. Instead of estimating $p(p-1)/ 2$ parameters, we could estimate $pk$ parameters with $1\leq k\leq p$. By directly estimating the latent vectors we have the possibility to explore several measures of dissimilarity available in \cite{Cox00}. This is topic of future research.

\begin{appendix}
\renewcommand\thesection{Appendix A}


\section[]{}

\subsection{Covariance functions \label{Covs}}

The univariate covariance function used in Section \ref{application} is given by
\begin{equation}\label{UNI}
C(\textbf{s})=\sigma^{2} \left(1+\left(\frac{h}{\phi}\right)^{2}\right)^{-1}, \nonumber
\end{equation}
with $h=\|\textbf{s}-\textbf{s}'\|$, $\textbf{s}$, $\textbf{s}' \in D$, $\sigma^{2}$ the variance of the variable and $\phi$ the spatial range.

\noindent The separable multivariate covariance function used in Section \ref{application} is given by
\begin{equation}\label{SEP}
C_{ij}(\textbf{s})=a_{ij} \left(1+\left(\frac{h}{\phi}\right)^{2}\right)^{-1}, \hspace{0.5cm} i,j=1,...p. \nonumber
\end{equation}
with $h=\|\textbf{s}-\textbf{s}'\|$, $\textbf{s}$, $\textbf{s}' \in D$, $a_{ij}$ the covariances of the components and $\phi$ the spatial range.

\subsection{Model comparison measures \label{Comp}}

As follows we present some measures considered for model comparison in the illustrations of our proposal.
\begin{enumerate}
\item Interval Score (IS) is given by
$$IS_{\alpha}(l,u;x) = (u-l)+\frac{2}{\alpha}(l-x)\mathtt{I}_{[x<l]}+\frac{2}{\alpha}(x-u)\mathtt{I}_{[x>u]}$$
where $l$ and $u$ represent for the forecaster quoted $\frac{\alpha}{2}$ and $1-\frac{\alpha}{2}$ quantiles. According to \cite{GneitRaf07}, the forecaster is rewarded for narrow prediction intervals, and he or she incurs a penalty, the size of which depends on $\alpha$, if the observation misses the interval.

\item Logarithm of the Pseudo Marginal Likelihood (LPML) is a cross-validation with log likelihood as the criteria,
$$ LPML = \sum_{i=1}^{n} log(CPO_i)$$
where $CPO_{i}$ is the Conditional Predictive Ordinate.
According to \cite{Ibrahim01b}, for the $i^{th}$ observation, the CPO statistic is defined as
$$
CPO_i=p(\textbf{y}_i|\textbf{y}_{(-i)})=\int p(\textbf{y}_i|\boldsymbol{\theta},\textbf{x}_i) \pi(\boldsymbol{\theta}|\textbf{y}_{(-i)}) d\boldsymbol{\theta}
$$
where $\textbf{y}_i$ denotes the response variable and $\textbf{x}_i$ is the vector of covariates for case $i$, $\textbf{y}_{(-i)}$ denotes the data without $\textbf{y}_i$, and $\pi(\boldsymbol{\theta}|\textbf{y}_{(-i)})$ is the posterior density of $\boldsymbol{\theta}$ based on the data $\textbf{y}_{(-i)}$.
Following \cite{Lesaffre12}, we are interested in computing the $CPO$ using MCMC output, so a simple derivation shows how to compute $CPO_i$:
$$
\frac{1}{p(\textbf{y}_i|\textbf{y}_{(-i)})} = \frac{p(\textbf{y}_{(-i)})}{p(\textbf{y})}= \int \frac{p(\textbf{y}_{(-i)} | \boldsymbol{\theta}) p(\boldsymbol{\theta})}{p(\textbf{y})} d\boldsymbol{\theta} = \int \frac{1}{p(\textbf{y}_i| \boldsymbol{\theta})} p(\boldsymbol{\theta}|\textbf{y}) d\boldsymbol{\theta} = E_{\boldsymbol{\theta}|\textbf{y}}\left(\frac{1}{p(\textbf{y}_i| \boldsymbol{\theta})} \right)
$$
where $\textbf{y} = (\textbf{y}_i,\textbf{y}_{(-i)})^{T}$. This derivation makes use of the conditional independence of the $\textbf{y}_i$ given $\boldsymbol{\theta}$. Then, we can estimate $CPO_i$ as an harmonic mean, that is,
$$
\widehat{CPO}_i = \left(\frac{1}{K} \sum_k \frac{1}{p(\textbf{y}_i| \boldsymbol{\theta}^{(k)})} \right)^{-1}.
$$

\end{enumerate}

\end{appendix}

\baselineskip=12pt
\paragraph{Acknowledgments} 
\

\noindent This research was performed while Rafael S. Erbisti was a PhD student at Federal University of Rio de Janeiro and the work of Rafael S. Erbisti was supported by CAPES. The work of Thais C. O. Fonseca was partially supported by the CNPq Grant PQ-2013 number 311441/2013-0. The work of Mariane B. Alves was partially supported by the CNPq Grant number 442608/2014-4.

\clearpage\pagebreak\newpage

\baselineskip=14pt \vskip 4mm\noindent

\bibliographystyle{plainnat}
\bibliography{ref}

\end{document}